\documentclass[journal=jctcce,manuscript=article]{achemso}
\usepackage{amssymb,amsmath,amsfonts,multicol,multirow,longtable,array,mathpazo}
\usepackage{lscape}
\usepackage[version=3]{mhchem} % Formula subscripts using \ce{}
\usepackage{appendix}
\usepackage{soul} %text highlighting
\usepackage[usenames]{color}
\usepackage{makecell}
\usepackage{enumerate}
\usepackage{xr}
\usepackage[T1]{fontenc} % Use modern font encodings
\usepackage{booktabs}
\usepackage{tikz}
\usepackage{threeparttable}
\usepackage{graphicx}
\usepackage{subfigure}
\usepackage{colortbl}
\usepackage{framed}
\usepackage{verbatim}
\usepackage{amsbsy}
\usepackage{bm}% bold math
\usepackage{bbm}
\usepackage[normalem]{ulem}
\usepackage{float}
\usepackage[linesnumbered,boxed]{algorithm2e}
\usepackage{algorithm2e}

\newcounter{xscheme}

\newfloat{Algorithm}{htbp}{alg}
\floatname{Algorithm}{Algorithm}

\newcounter{exe}[figure]
\newcommand{\iexe}{\refstepcounter{exe}\the\value{exe}:}

\setkeys{acs}{maxauthors = 0} % references of many authors

\author{Ning Zhang}
\affiliation{Beijing National Laboratory for Molecular Sciences, Institute of Theoretical and Computational Chemistry,
College of Chemistry and Molecular Engineering, Peking University, Beijing 100871, China}
\author{Wenjian Liu}\email{liuwj@sdu.edu.cn}
\affiliation{Qingdao Institute for Theoretical and Computational Sciences,
Shandong University, Qingdao, Shandong 266237, China}
\author{Mark R. Hoffmann}%\email{mark.hoffmann@email.und.edu}
\affiliation{Chemistry Department, University of North Dakota, Grand Forks, ND 58202-9024, U.S.A.}

\title{Further Development of iCIPT2 for Strongly Correlated Electrons}

\setkeys{acs}{articletitle=true}

\begin{document}
%\twocolumn[\begin{@twocolumnfalse}
%\end{@twocolumnfalse}]

\begin{abstract}
The efficiency of the recently proposed iCIPT2 [iterative configuration interaction (iCI) with selection and second-order perturbation theory (PT2); J. Chem. Theory Comput. 16, 2296 (2020)]
for strongly correlated electrons is further enhanced
(by up to $20\times$)
by using (1) a new ranking criterion for configuration selection, (2) a new particle-hole algorithm for Hamiltonian construction over
randomly selected configuration state functions (CSF), and (3) a new data structure for the quick sorting of the joint variational and first-order interaction spaces. Meanwhile, the memory requirement is also reduced greatly. As a result, this improved implementation of iCIPT2 can handle one order of magnitude more CSFs than the previous version, as revealed by taking the chromium dimer and an iron-sulfur cluster, \ce{[Fe2S2(SCH3)]4}$^{2-}$,
as examples.
\end{abstract}

\maketitle

\clearpage
\newpage

\section{Introduction}
As pointed out recently\cite{LiuPerspective2020,LiuSciChina2020},
among the three components of electronic structure theory (i.e., relativity, correlation, and QED), it is correlation that
is most challenging, especially for strongly correlated systems.
Roughly speaking, a strongly correlated system features multiple open-shell orbitals or nearly degenerate electronic states thanks to the existence of a dense set of energetically adjacent frontier
orbitals, such that there does not exist a single, leading component in the wave functions. Because of this, single-reference methods are bound to fail, especially for
those low-spin states. Instead, multi-reference methods\cite{MRPT2Rev1,MRPT2Rev2,MRCCRev,MRCIRev,MRCIrev2012,MRCIrev2018} are needed even for a qualitative description of such states.
A unanimous assumption underlying standard multi-reference methods is that the full many-electron
Hilbert space can be decomposed into a model space $P$ and the remaining complement $Q=1-P$, which are responsible for the static and dynamic components of the overall correlation,
respectively. Then, according to when the static and dynamic correlations are handled, such methods can be classified into three families, viz.
``static-then-dynamic'', ``dynamic-then-static'', and ``static-dynamic-static'' (SDS)\cite{iCI}.
 Given their great success in investigating low-lying electronic states
of many chemical systems, it must be realized that such multi-reference methods still have strong limitations. For instances,
it is by no means trivial to maintain the same model space, so as to produce smooth potential energy surfaces. Even if this can be achieved,
there is no guarantee that the chosen model space is equally good for all geometries and all target states. Even more seriously, those states
that have little projections in the chosen model space cannot well be described. Taking all these facts into account, one can classify
chemical systems into weakly correlated (which can well be described by single-reference methods), moderately correlated (which can well be described
by standard multi-reference methods with a moderate model space, typically much smaller than the complete active space (CAS) of 18 electrons in 18 orbitals),
and strongly correlated (which require a large model space and are therefore
beyond the capability of standard multi-reference methods).
As a matter of fact, the concept of model space looses its original meaning for the so-defined strongly correlated systems
because the states in a large model space $P$ and those formally in the orthogonal complement $Q$ are heavily intersected in both energy and composition.
In other words, the static and dynamic correlations are in such cases strongly entangled and interchangeable.
Moreover, solving the large-model-space problem itself is already a heavy task, needless to say
the subsequent treatment of dynamic correlation (NB: the larger the model space, the larger the first-order interacting space (FOIS)).
So the question is how to design an adaptive method that can adapt to the variable static correlation automatically and can meanwhile handle the residual
dynamic correlation efficiently. It appears that the most promising way to this end is to introduce some selection procedure,
so as to build up progressively a compact yet accurate variational space $P$,
in full accordance with the nature of the system under concern. In this spirit, a number of near-exact approaches have been designed in the past, including
density matrix renormalization group (DMRG)\cite{DMRG1992,DMRG1993,DMRG1999,DMRG2001,DMRG2002,DMRGrev2011,SA-DMRG2012,DMRG2014,DMRG2015,spDMRG2017,DMRGrev2020},
full configuration interaction (FCI) quantum Monte Carlo (FCIQMC)\cite{FCIQMC2009,i-FCIQMC,FCIQMC2017excited,S-FCIQMC,holmes2016efficient,FCIQMC-Ten-no,AS-FCIQMC2019,sCI-i-FCIQMC,FCIQMCGUGA,QN-FCIQMC2020},
cluster-analysis-driven FCIQMC\cite{CAD-FCIQMC2018}, adaptive coupled-cluster (CC)\cite{AdaptiveCC}, full CC reduction\cite{FCCR2018,FCCRPT2},
deterministically projected FCI\cite{ProjCI},
many-body expanded FCI\cite{MBE-FCI2017,MBE-FCI2018,MBE-FCI2019,MBE-FCI2019generalized}, incremental FCI\cite{iFCI2017a,iFCI2017b}, rank-reduction FCI\cite{RR-FCI},
fast randomized interaction-based FCI\cite{FRI-FCI}, intrinsic scaling-based correlation expansion\cite{IntrinsicScaling1,IntrinsicScaling2}, as well as selected CI (sCI) \cite{MCCI1998,MCCIPT2,MCCI2017,ML-CI,HBCI2016,HBCI2017a,HBCI2017b,HBCI2017c,HBCI2018,HBCI2020,LambdaCI2014,ACI2016,ACI2017,ACI2018,ASCI2016,ASCI2017,ASCI2020,ASCI2018PT2,SDS,iCI,SDSPT2,iCIPT2,DressedSCI2018,CD-FCI,CIPSI-DMCd,CIPSI-DMCe}.
Among these, the sCI type of approaches, which are largely a revival of the very old ideas\cite{Davidson1969,Whitten1969,CIPSIa,CIPSIb,MRDCIb} in one way or another,
are the simplest in structure and rival those more sophisticated approaches as shown in a recent blind test\cite{BlindTest}.
As the newest member of the sCI family, the iCIPT2 approach\cite{iCIPT2} proposed by the present authors, i.e., the combination of iterative CI (iCI)\cite{iCI}
with selection and second-order perturbation theory (PT2) has a number of salient features:
\begin{enumerate}[(1)]
\item Born from the restricted SDS framework\cite{SDS} for strongly correlated electrons, iCI\cite{iCI} is a parameter-free, exact solver of the FCI eigenvalue problem.
It constructs and diagonalizes a very small Hamiltonian matrix $\bar{\mathbf{H}}$ (see Sec. \ref{SeciCI}) at each macro/micro-iteration but can converge quickly to the exact solutions.
The major bottleneck of iCI lies therefore only in the construction of the Hamiltonian matrix $\mathbf{H}$ in the basis of Slater determinants (SD) or configuration state functions (CSF) before contracted to
$\bar{\mathbf{H}}$ for diagonalization. Since $\mathbf{H}$ is extremely sparse, the combination of iCI with selection is a natural choice, leading to ``selected iCI'' (SiCI) as an near-exact solver of FCI.
To be more practical, the (one-step) SiCI can be performed in two steps, with the first step accounting for the static correlation whereas the second step for a second-order perturbative treatment of the dynamic correlation.
This can be achieved by a single parameter $C_{min}$, which controls the size of the variational space and hence the final accuracy. This scheme has been dubbed as iCIPT2\cite{iCIPT2}.
\item The selection is carried out over the entire Hilbert space or a very large active space (which can in this context be viewed as a coarse-grained selection step). Those unimportant doubly excited orbital configurations (CFG) are first screened out based on the upper bounds of the elements of $\mathbf{H}$.
After this integral-driven screening, the individual CSFs of a selected CFG are further selected using their approximate first-order coefficients. In short,
although the selection is performed on individual CSFs, it is CFGs that are used as the organizing units. Given a coefficient pruning-threshold $C_{min}$,
the selection of important CFGs/CSFs is performed iteratively until convergence.
\item At each iteration for the growth of wave function, the FOIS is decomposed into disjoint subspaces, so as to reduce memory requirement on one hand and facilitate parallelization on the other.
\item A new technique called TUGA (tabulated unitary group approach) is introduced to compute and reuse the basic coupling coefficients (BCC), i.e., the matrix elements of $U(n)$ generators over CSF pairs.
\item Because of the use of CSFs as the many-electron basis, full spin symmetry is always maintained, which is of vital importance for describing low-spin states of general open-shell systems as well as
subsequent perturbative treatment of spin-orbit couplings.
\item Upon termination of the selection, dynamic correlation is estimated by using the state-specific Epstein-Nesbet type of second-order perturbation theory (ENPT2).
\item The linear relationship between the total versus second-order correlation energies allows for an accurate extrapolation.
\end{enumerate}

The present contribution amounts to further enhancing the efficiency of iCIPT2 by using
(a) a new ranking criterion for the selection of CFGs/CSFs (see Sec. \ref{Selection}), (b) a new particle-hole algorithm for the construction and update of $\mathbf{H}$ in the variational space (see Sec. \ref{FastHmat}),
and (c) the Timsort algorithm\cite{Timsort,TimsortCode} for sorting the joint variational and first-order interacting spaces (see Sec. \ref{PT2}). Before these, we ought to outline the
SDS family of methods in Sec. \ref{SeciCI} and the essentials of TUGA in Sec. \ref{TUGA}. A close comparison of TUGA
with Table-CI\cite{TableCI1980,TableCI1986,TableCI1995,TableCIMark2} is also given in the latter, since they are both configuration-driven alogithms. The chromium dimer
and an iron-sulfur cluster, \ce{[Fe2S2(SCH3)]4}$^{2-}$,
are finally taken as examples to reveal the efficacy of the improved implementation of iCIPT2 (see Sec. \ref{Results}).
The presentation is closed with concluding remarks in Sec. \ref{Conclusion}.

\section{The SDS family of methods: SDSCI, SDSPT2, iCI, iCIPT2}\label{SeciCI}
The \emph{restricted} ``static-dynamic-static'' (SDS) framework\cite{SDS} for strongly correlated electrons gives rise first to the SDSCI approach, which
is a minimal MRCI with the following form
for the wave function $|\Psi_i\rangle$ of state $i\in[1, N_P]$,
\begin{align}
|\Psi_i\rangle&=\sum_{k=1}^{mN_P}|\tilde{\Phi}_k\rangle C_{k,i},\quad m=3 \mbox{ or } 4,\label{SDSCIW}\\
|\tilde{\Phi}_k\rangle&=|\Psi_k^{(0)}\rangle=\sum_{|J\nu\rangle\in P}|J\nu\rangle \bar{C}_{\nu,k}^{J(0)},\quad k\in[1,N_P],\\
|\tilde{\Phi}_{k+N_P}\rangle&=|\Psi_k^{(1)}\rangle=Q\frac{1}{E_k^{(0)}-H_0}QH|\Psi_k^{(0)}\rangle=\sum_{|J\nu\rangle\in Q}|J\nu\rangle\bar{C}_{\nu,k}^{J(1)},\quad  k\in[1, N_P],\label{1stW}\\
|\tilde{\Phi}_{k+2N_P}\rangle&=P_{\mathrm{s}}|\Theta_k\rangle,\quad k\in[1, N_P],\label{2ndW}\\
|\tilde{\Phi}_{k+3N_P}\rangle&=P_{\mathrm{s}}|\Theta_k^\prime\rangle,\quad k\in[1, N_P]\quad (\mbox{ if } m>3),\label{3rdW}\\
P&=\sum_{J\nu}^{d_R}|J\nu\rangle\langle J\nu|=P_{\mathrm{m}}+P_{\mathrm{s}},\quad P_{\mathrm{m}}=\sum_{k=1}^{N_P}|\Psi_k^{(0)}\rangle\langle\Psi_k^{(0)}|,\quad Q=1-P.
\end{align}
Here, the first two sets of $N_P$ functions $\{|\tilde{\Phi}_k\rangle\}$ are nothing but the zeroth-order (primary)
and first-order (external) functions, respectively. To account for changes in the static correlation (described by $\{|\Psi_k^{(0)}\rangle\}_{k=1}^{N_P}$) due to the presence
of dynamic correlation (described by $\{|\Psi_k^{(1)}\rangle\}_{k=1}^{N_P}$), $N_P$ secondary (buffer) functions $\{|\Theta_k\rangle\}_{k=1}^{N_P}$, Eq. \eqref{2ndW}, are further introduced, which can be defined in
a number ways. For instance, the \textit{not-energy-biased} Lanczos-type functions (denoted as Buf(1)), $\{H|\Psi_k^{(1)}\rangle\}_{k=1}^{N_P}$, turn out to be very effective\cite{SDS}.
An alternative choice might be the second set of $N_P$ zeroth-order functions $\{|\Psi_k^{(0)}\rangle\}_{k=N_P+1}^{2N_P}$ (denoted as Buf(0)) in the spirit of intermediate Hamiltonian theory, the first set of $N_P$ second-order functions $\{|\Psi_k^{(2)}\rangle\}_{k=1}^{N_P}$) (denoted as Buf(2)) in the spirit of perturbation theory or the $N_P$ approximate eigenvectors of the preceding iteration (denoted as Buf(3)) in the spirit of conjugate gradient theory\cite{iVI}.
While Buf(2) is very expensive, both Buf(0) and Buf(3) are very easy to handle because the Hamiltonian matrix elements can readily be evaluated\cite{iVI}. As a matter of fact,
Buf(0) and Buf(3) can be applied simultaneously without increasing the computational overhead. In this case, we will have $m=4$ in Eq. \eqref{SDSCIW}. Otherwise, we have $m=3$ in Eq. \eqref{SDSCIW} with Buf(1) alone.
The yet unknown expansion coefficients are to be determined by the generalized secular equation
\begin{eqnarray}
\bar{\mathbf{H}}\mathbf{C}=\mathbf{S}\mathbf{C}\mathbf{E}. \label{EigenEQ}
\end{eqnarray}
Since all the functions $\{|\tilde{\Phi}_k\rangle\}_{k=N_P+1}^{mN_P}$ are fully contracted (i.e., linear combinations of the CSFs $\{|\nu\rangle\}$ generated from CFGs $\{|J\rangle\}$)
and are specific to the primary state $|\Psi_k^{(0)}\rangle$,
the dimension of Eq. \eqref{EigenEQ} is just 3 (or 4) times the number ($N_P$) of target states, no matter how many electrons and how many orbitals are correlated.
Although restricted as such, it has been shown\cite{SDSCI} that SDSCI is a very effective variational method for both ground and excited states.
Two extensions of SDSCI have been considered so far, SDSPT2\cite{SDS,SDSPT2} and iCI\cite{iCI}, both with Buf(1) alone. The former amounts to replacing the $QHQ$ block of the Hamiltonian matrix in Eq. \eqref{EigenEQ} with $QH_0Q$.
Different from most variants of multi-reference second-order perturbation theory (MRPT2), the CI-like SDSPT2 treats single and multiple states in the same way and is particularly advantageous when a number of states are nearly degenerate,
thanks to the efficacy of the secondary
states in revising the coefficients of the primary states. In contrast, iCI takes the solutions of Eq. \eqref{EigenEQ} as new primary states $\{|\Psi_k^{(0)}\rangle\}_{k=1}^{N_P}$ and
repeats the SDS procedure \eqref{SDSCIW} until convergence. Clearly,
each iteration (defined as macro-iteration) accesses a space that
is higher by two ranks than that of the preceding iteration.
Up to $2i$-tuple excitations (relative to the initial primary
space) can be accessed if $i$ macro-iterations are carried out.
A few micro-iterations $j$ can be invoked at
each macro-iteration so as to relax the contraction coefficients.
As such, every iCI($i, j$) corresponds to a physically meaningful model.
It has been shown both theoretically and numerically that iCI can converge quickly from above to FCI even when starting with a very poor initial guess.
More generally, iCI can be viewed as a particular sequential, exact partial diagonalization of a huge matrix, by getting first the roots of one portion
of the matrix and then those of an enlarged portion, until the full matrix,
whereas its micro-iterations can be generalized to an iterative vector interaction (iVI) approach\cite{iVI}
for the roots of a given matrix treated as a whole. Again, each iteration of iVI constructs and diagonalizes a $mN_P\times mN_P$ matrix ($m=3 \mbox{ or } 4$) in the form of Eq. \eqref{EigenEQ}.
In particular, by combining with the energy-vector following technique,
iVI can directly access interior roots belonging to a predefined window, without knowing the number and characters of the roots\cite{iVI-TDDFT}.

As stated before, iCIPT2 amounts to splitting iCI into two steps, the first of which accounts for the static correlation by selecting important CFGs/CSFs in an iterative manner,
whereas the second of which accounts for the residual dynamic correlation via a second-order perturbation theory. It is the improvement of iCIPT2 that is under concern here.

\section{TUGA}\label{TUGA}
The very first issue pertinent to iCIPT2 is how to evaluate the Hamiltonian matrix elements over randomly selected CSFs $\{|J\nu\rangle\}$.
In this context, the configuration-driven Table-CI approach\cite{TableCI1980,TableCI1986,TableCI1995,TableCIMark2} may be a possible choice. However,
it involves line-up permutations of open-shell orbitals, which become ineffective for CFGs of many open-shell orbitals (say, $>10$).
To avoid such line-up permutations, we have introduced a ``tabulated unitary group approach'' (TUGA)\cite{iCIPT2} (originally called
``tabulated orbital configuration-based unitary group approach''). Since the UGA\cite{CSF1,Paldus1980,CSF2}
is itself rather involved, we here recapitulate the essentials of TUGA. The first step is to break
the spin-free, second-quantized Hamiltonian
\begin{eqnarray}
H&=&\sum_{i,j}h_{ij}E_{ij}+\frac{1}{2}\sum_{i,j,k,l}(ij|kl)e_{ij,kl},\label{sfH}\\
E_{ij}&=&\sum_{\sigma}a_{i\sigma}^\dag a_{j\sigma}=E_{ji}^\dag,\label{EijOp}\\
e_{ij,kl}&=&\sum_{\sigma,\tau} a_{i\sigma}^\dag a_{k\tau}^\dag a_{l\tau} a_{j\sigma}=E_{ij}E_{kl}-\delta_{jk}E_{il}=\{E_{ij}E_{kl}\}\nonumber\\
&=&\{E_{kl}E_{ij}\}=e_{kl,ij}=e^\dag_{ji,lk}\label{eijkl}
\end{eqnarray}
into a form that is consistent with the diagrams employed in the UGA\cite{Paldus1980} for
the BCC between CSFs. Specifically,
\begin{eqnarray}
H&=&H_1^0+H_2^0 + H_1^1 + H_2^1 + H_2^2,\\
H_1^0&=&\sum_i h_{ii} E_{ii},\label{H1-0}\\
H_1^1&=&\sum_{i<j} h_{ij}E_{ij}+\sum_{i>j}h_{ij}E_{ij},\label{H1-1}\\
H_2^0&=&\frac{1}{2}\sum_i(ii|ii)E_{ii}(E_{ii}-1)+\sum_{i<j}\left[(ii|jj)E_{ii}E_{jj}+(ij|ji)e_{ij,ji}\right],\label{H2-0b}\\
H_2^1&=&\sum_{i\ne j}\left[(ii|ij)(E_{ii}-1)E_{ij}+(ij|jj)E_{ij}(E_{jj}-1)\right]\nonumber\\
&+&\sum_{i\ne j\ne k}\left[(ij|kk)E_{ij}E_{kk}+(ik|kj)e_{ik,kj}\right],\label{H2-1b}\\
H_2^2&=&\sum_{i\le k}\sum_{j\le l}^\prime \left[ 2^{-\delta_{ik}\delta_{jl}}(ij|kl)e_{ij,kl}+(1-\delta_{ik})(1-\delta_{jl})(il|kj)e_{il,kj}\right],\label{EqnDiff2}
\end{eqnarray}
where the superscripts 0, 1, and 2 in $H_i$  ($i = 1, 2$) indicate that the terms contribute to the
Hamiltonian matrix elements over two CFGs that are related
by zero, single, and double excitations, respectively, whereas the prime in Eq. \eqref{EqnDiff2} indicates that $\{j, l\}\cap\{ i, k\} =\emptyset$.
For two identical CFGs $|I\rangle$, the Hamiltonian matrix elements can be calculated most efficiently as\cite{iCIPT2}
\begin{align}
\langle I\mu|H_1^0+H_2^0|I\nu\rangle&=\delta_{\mu\nu}\{\frac{1}{2}\sum_i n_i^R[h_{ii}+\epsilon_i+(n_i^R-2)g_{ii}]+\sum_i\Delta_i^I[\epsilon_i+(n_i^R-1)g_{ii}]\nonumber\\
&+\sum_{i\le j}\Delta_i^Ig_{ij}\Delta_j^I\}+\sum_{i<j}(ij|ji)\langle I\mu|e_{ij,ji}^1|I\nu\rangle \delta(n_i^I, 1)\delta(n_j^I,1),\label{H-diag-3}\\
f_{ij}&=h_{ij}+\sum_k n_k^R[(ij|kk)-\frac{1}{2}(ik|kj)],\label{Inter1}\\
\epsilon_i&=f_{ii}=h_{ii}+\sum_k n_k^R g_{ik},\quad g_{ik}=(ii|kk)-\frac{1}{2}(ik|ki)=g_{ki},\label{Inter2}\\
\Delta_i^I&=n_i^I-n_i^R,
\end{align}
where $\{n_i^R\}$ are the occupation numbers of the spatial orbitals of a common reference CFG $|R\rangle$ (either Hartree-Fock (HF) or restricted open-shell HF).
The superscript 0 or 1 in $e_{ij,kl}$ (see Eq. \eqref{H-diag-3}) refers to the intermediate angular momentum $X$ due to the coupling of two spin-$\frac{1}{2}$ functions.
If $|I\rangle=E_{ij}|J\rangle$, the corresponding Hamiltonian matrix elements read (under the Yamanouchi-Kotani (YK) phase\cite{Paldus1980})
\begin{equation}
\begin{split}
\langle I\mu|H_1^1+H_2^1|J\nu\rangle&=\left\langle I\mu|E_{i j}|J\nu\right\rangle\left[f_{ij}+\sum_k\Delta_k^J[(ij|kk)-\frac{1}{2}(ik|kj)]\right.\\
&+\left.\frac{1}{2}n_i^J(ii|ij)+(\frac{1}{2}n_j^J-1)(ij|jj)\right]\\
&+\sum_{k\in\mbox{ exterior open}}(ik|kj)\langle I\mu| e_{ik,kj}^1|J\nu\rangle\\
&+\sum_{k\in\mbox{ interior open}}(ik|kj)\left[\frac{1}{2}\left\langle I\mu|E_{i j}|J\nu\right\rangle+\langle I\mu| E_{kj}E_{ik}|J\nu\rangle\right],
\end{split}\label{EqnDiff1}
\end{equation}
where `exterior open' and `interior open' emphasize that level $k$ belongs to
the non-overlapping ($k<\min(i, j)$ or $k>\max(i,j)$) and overlapping ($\min(i,j)<k<\max(i,j)$) cases, respectively, and is singly occupied.
On the other hand, if $|I\rangle=e_{ij,kl}|J\rangle$ subject to $\{j, l\}\cap\{ i, k\} =\emptyset$, the Hamiltonian matrix elements read
\begin{eqnarray}
\langle I\mu|H_2^2|J\nu\rangle=\left[ 2^{-\delta_{ik}\delta_{jl}}(ij|kl)\langle I\mu|e_{ij,kl}|J\nu\rangle+(1-\delta_{ik})(1-\delta_{jl})(il|kj)\langle I\mu|e_{il,kj}|J\nu\rangle\right].\label{DoubleMat}
\end{eqnarray}

At this stage two important points can be observed:
\begin{enumerate}[(1)]
\item Thanks to the conjugacy (bra-ket inversion) relations $\langle I\mu|E_{ij}|J\nu\rangle=\langle I\mu|E_{ij}|J\nu\rangle^*=\langle J\nu|E_{ji}|I\mu\rangle$ and $\langle I\mu|e_{ij,kl}|J\nu\rangle=\langle I\mu|e_{ij,kl}|J\nu\rangle^*=\langle J\nu|e_{ji,lk}|I\mu\rangle$ in the absence of spin-orbit couplings,
only generators $E_{ij}$ with $i>j$ and $e_{ij,kl}$ with $k\ge i$ and $k>l\ge j$ need to be considered explicitly. They correspond to the s2, c$x$ and d$x$ ($x\in[1,7]$) types of diagrams
shown in Figs. 2 and 3 of Ref. \citenum{iCIPT2}. More specifically, the s2, c4, c6 and d2 diagrams are required for the evaluation of Eq. \eqref{EqnDiff1}, while the other c$x$ and d$x$ types of diagrams
are needed for the evaluation of Eq. \eqref{DoubleMat}. Once such BCCs are available, those conjugate ones can be obtained simply by matrix transpose.
\item The BCCs $\langle I\mu|E_{ij}|J\nu\rangle$ and $\langle I\mu|e_{ij,kl}|J\nu\rangle$
depend only on the relative occupations of the CFG pair $(I, J)$ but not on the individual orbitals $\{i, j, k, l\}$.
Moreover, the common zero or doubly occupied orbitals need not be considered explicitly because their segment values are just one under the YK phase.
Therefore, they can be rewritten as $\langle I\mu|E_{\bar{i}\bar{j}}|J\nu\rangle$ and $\langle I\mu|e_{\bar{i}\bar{j},\bar{k}\bar{l}}|J\nu\rangle$, respectively, in terms of
the reduced orbital indices (ROI) $\{\bar{i}, \bar{j}, \bar{k},\bar{l}\}$ after deleting the common zero or doubly occupied orbitals.
Note that the ROIs have one-to-one correspondence with the original orbital indices
(see the example in Table \ref{CoeffExam}). This way, the same BCCs can be used for very many integrals sharing the same ROIs.
\end{enumerate}

After deleting the common zero or doubly occupied orbitals in the bra and ket CFGs, we are left with seven occupation patterns, each of which can be assigned a code number
as shown in Table \ref{ROT}. Every pair of CFGs can therefore be characterized by a reduced occupation table (ROT) consisting of two sequences, ROT\_Orb and ROT\_Code. The former records the orbital indices to fetch molecular integrals, whereas the latter records the corresponding code sequence (e.g., (012430) for the example shown in Table \ref{CoeffExam}) to determine the generators, with the following rules:
(1) if CFG $|I\rangle$ arises from $|J\rangle$ by exciting one electron from orbital $j$ to orbital $i$, $\bar{j}$ would then correspond to code 2 or 4, while $\bar{i}$ to code 1 or 3 because
of the relations $n_j^I=n_j^J-1$ and $n_i^I=n_i^J+1$. (2) If CFG $|I\rangle$ arises from $|J\rangle$ by exciting two electrons from orbitals $j$ and $l$ $(\ge j)$ to orbitals $i$ and $k$ $(\ge i)$ [NB: $\{i,k\}\cap\{j,l\}=\emptyset$],
$\bar{j}$ and $\bar{l}$ $(\ge \bar{j})$ would correspond to code 2, 4 or 6, while $\bar{i}$ and $\bar{k}$ $(\ge \bar{i})$ to code 1, 3 or 5 because of the relations $n_j^I=n_j^J-1$, $n_l^I=n_l^J-1$, $n_i^I=n_i^J+1$ and $n_k^I=n_k^J+1$. Codes 6 and 5 further imply $\bar{j}=\bar{l}$ and $\bar{i}=\bar{k}$, respectively. (3) If $\bar{i}<\bar{j}$ in $E_{\bar{i}\bar{j}}$ or $\bar{k}<\bar{l}$ in $e_{\bar{i}\bar{j},\bar{k}\bar{l}}$, a bra-ket inversion should be invoked when calculating
the BCCs in terms of the diagrams documented in Table \ref{PaldusType}. To search and sort the code sequences efficiently,
each ROT\_Code will be converted to an array of 64-bit integers (3 bits per code). It is clear that, to reutilize the BCCs efficiently, the CFG pairs
with the same ROT\_Code must be grouped together. Instead of the red-black tree type of
bilinear search employed before\cite{iCIPT2}, we adopt here an array-based sorting algorithm:
(a) generate all connected CFG pairs and store them in one array; (b) loop over each CFG pair and identify its ROT\_Orb
and ROT\_Code; (c) sort the array based on ROT\_Code. This way, those CFG pairs sharing the same ROT\_Code (and hence the same BCCs) are adjacent to each other and form a segment in the array.
To expedite the sorting process,
%the array can be classified into $N$ disjoint subarrays in advance.
%In the best case, the number of comparison can be reduced to $\frac{N_{pair}}{N}\log(\frac{N_{pair}}{N})\times N = N_{pair}\log(N_{pair})-N_{pair}\log(N)$
%with $N_{pair}$ being the number of CFG pairs.
the code sequences can further be characterized by their lengths and category numbers (CN) defined in Table \ref{PaldusType}.
The latter just number different ranges of the generators ($\mathrm{CN}=0$ for singles and $\mathrm{CN}\in[1,7]$ for doubles) and are hence directly related to the diagrams required for the evaluation of the BCCs.
This new sorting algorithm is significantly faster than the red-black trees employed before, although the latter is not really expensive for this purpose.

 \begin{table}[!htp]
	\centering
	\caption{Illustration on the reduced orbital indices (ROI)}
\begin{threeparttable}	\begin{tabular}{p{4cm}p{1cm}<{\centering}p{1cm}<{\centering}p{1cm}<{\centering}p{1cm}<{\centering}p{1cm}<{\centering}p{1cm}<{\centering}p{1cm}<{\centering}p{1cm}<{\centering}p{1cm}<{\centering}}\toprule
		OrbIndx&0&1&2&3&4&5&6&7\\
		bra occ&1&0&1&0&2&1&2&1\\
		ket occ&1&0&0&1&2&2&1&1\\
        ROT\_Orb&0&$\times$&2&3&$\times$&5&6&7\\
		ROI&$\bar{0}$&$\times$&$\bar{1}$&$\bar{2}$&$\times$&$\bar{3}$&$\bar{4}$&$\bar{5}$\\
		ROT\_Code\tnote{*}&0&$\times$&1&2&$\times$&4&3&0\\\bottomrule
	\end{tabular}
		\begin{tablenotes}
		\item[*] See Table \ref{ROT}.
		\end{tablenotes}
\end{threeparttable}
	\label{CoeffExam}
\end{table}

\begin{table}[!htp]
	\centering
	\caption{Code numbers for the relative occupation patterns of CFG pairs}
	\begin{tabular}{p{1.5cm}p{1cm}<{\centering}p{1cm}<{\centering}p{1cm}<{\centering}p{1cm}
			<{\centering}p{1cm}<{\centering}p{1cm}<{\centering}p{1cm}<{\centering}p{1cm}
			<{\centering}p{1cm}<{\centering}}\toprule
		bra occ&1&1&0&2&1&2&0&0&2\\
		ket occ&1&0&1&1&2&0&2&0&2\\
		code   &0&1&2&3&4&5&6&$\times$&$\times$\\\bottomrule
	\end{tabular}\label{ROT}
\end{table}

\begin{table}
	\centering
	\caption{Correspondence between UGA diagrams and category numbers (CN) of code sequences recorded in ROT\_code}
	\begin{threeparttable}
		\begin{tabular}{cccc|c}\toprule
			CN              & Generator                     & Range                                   & Diagram\tnote{a} & $\tilde{H}^{IJ}$ \tnote{b}\\ \toprule
			\multirow{4}{*}{0}           & $E_{ij}$                      & $i>j$                      & s2 &\\
			                             & \multirow{3}{*}{$e_{ik,kj}^1$}& $i>j>k$                    & c4 &\\
			                             &                               & $k>i>j$                    & c6 &\\
			                             &                               & $i>k>j$                    & d2 &\\
%			\midrule
			\cline{1-5}
			\multirow{2}{*}{1}           & $e_{ij,kl}$                   & \multirow{2}{*}{$k>l>j>i$} & c1 &\multirow{6}{*}{\makecell{$\max(|(ij|kl)+(il|kl)|$,\\$\sqrt{3}|(ij|kl)-(il|kj)|)$}}\\
			                             & $e_{il,kj}$                   &                            & c3 &\\
              \cline{1-4}
			\multirow{2}{*}{2}           & $e_{ij,kl}$                   & \multirow{2}{*}{$k>l>i>j$} & d1 &\\
			                             & $e_{il,kj}$                   &                            & c5 &\\
              \cline{1-4}
			\multirow{2}{*}{3}           & $e_{ij,kl}$                   & \multirow{2}{*}{$k>i>l>j$} & d3 &\\
			                             & $e_{il,kj}$                   &                            & d5 &\\
              \cline{1-5}
			4                            & $e_{ij,kj}$                   & $k>j>i$                    & c2 &\multirow{2}{*}{2$|(ij|kj)|$}\\
			5                            & $e_{ij,kj}$                   & $k>i>j$                    & d4 &\\
              \cline{1-5}
			6                            & $e_{il,ij}$                   & $i>l>j$                    & d6 &2$|(il|ij)|$\\		
              \cline{1-5}
			7                            & $e_{ij,ij}$                   & $i>j$                      & d7 &$|(ij|ij)|$\\	
          \bottomrule
		\end{tabular}
		\begin{tablenotes}
        \item[a] See Figs. 1-3 in Ref. \citenum{iCIPT2}.
		\item[b] Estimates of $|\langle I\mu|H_2^2|J\nu\rangle|$, which are upper bounds\cite{iCIPT2} for $\mathrm{CN}\in [4,7]$, exact for
          $\mathrm{CN}\in [1,3]$ if $|I\rangle$ or $|J\rangle$ is a closed shell CFG, but not guaranteed to be upper bounds for $\mathrm{CN}\in [1,3]$
          if both $|I\rangle$ and $|J\rangle$ are open-shell CFGs (NB: slight deviations from the upper bounds can be
          taken care of by adjusting $C_{min}$).
		\end{tablenotes}
	\end{threeparttable}
\label{PaldusType}
\end{table}

To be compatible with the desired generator ranges, we have to represent and store properly the CFGs and corresponding CSFs.
A CFG can first be represented by an array of occupation numbers, Occ[$i$]=$n_i$, $i\in [0, N_{orb}-1]$, with $N_{orb}$ being the total number of spatial orbitals.
The comparison of a CFG pair can then be performed efficiently with Algorithm 3 presented previously\cite{iCIPT2}.
A CFG pair $(I, J)$ is said to be singly or doubly connected if $|I\rangle$ can be obtained by exciting one or two electrons from $|J\rangle$.
CFG $|I\rangle$ is stored preceding CFG $|J\rangle$ (i.e., $I<J$) if and only if $n_p^I>n_p^J$ for $p=\max\{x|n_x^I\neq n_x^J\}$.
A more compact representation of CFGs is to use two bits to store the occupancy number of each spatial orbital. Specifically, $(00)_2$, $(01)_2$ and $(11)_2$ are
used for $n_i =0$, 1, and 2, respectively. A CFG can then be represented by an array, OrbOccBinary, of 64-bit integers.
This representation was used for both bra and ket CFGs in our previous implementation. Now it is discovered that the single occupancy
in the bra CFG can be represented by $(10)_2$ instead of $(01)_2$, so as to make the identification of ROT\_Code more efficiently.
As can be seen from Table \ref{XOR}, the bitwise XOR operations on the so-defined binary occupation codes (BOC) always give $(00)_2$ for common zero or doubly occupied orbitals,
such that the identification of ROT\_Code can be achieved based fully on bit operations. Such double representation requires
$M_{CFG}=2\times 8\left(\lfloor \frac{N_{orb}-1}{32}\rfloor+1\right)$ bytes of memory for storing a CFG.

\begin{table}[!htp]
	\caption{Bitwise XOR operations on binary occupation codes (BOC)}
	\begin{tabular}{ccccc}\toprule
		Bra Occ & BOC & Ket Occ & BOC & XOR \\\toprule
		0 & $(00)_2$ & 0 & $(00)_2$ & $(00)_2$ \\
		2 & $(11)_2$ & 2 & $(11)_2$ & $(00)_2$ \\
		1 & $(10)_2$ & 1 & $(01)_2$ & $(11)_2$ \\
		0 & $(00)_2$ & 1 & $(01)_2$ & $(01)_2$ \\
		0 & $(00)_2$ & 2 & $(11)_2$ & $(11)_2$ \\
		1 & $(10)_2$ & 0 & $(00)_2$ & $(10)_2$ \\
		1 & $(10)_2$ & 2 & $(11)_2$ & $(01)_2$ \\
		2 & $(11)_2$ & 0 & $(00)_2$ & $(11)_2$ \\
		2 & $(11)_2$ & 1 & $(01)_2$ & $(10)_2$ \\ \bottomrule
	\end{tabular}
\label{XOR}
\end{table}

A CFG $|I\rangle$ can generate $N^I_{S,S}$, genealogically coupled CSFs $\{|I\mu\rangle\}_{\mu=0}^{N_{S,S}^I-1}$ of total spin $S$, with $N^I_{S,S}$ being
\begin{equation}
	N_{S,M=S}^I= \frac{2S+1}{\frac{1}{2}N_o^I+S+1}C_{N_o^I}^{\frac{1}{2}N_o^I-S},\label{NSS}
\end{equation}
where $N_o^I$ is the (seniority) number of open-shell orbitals in CFG $|I\rangle$.
Such CSFs are characterized uniquely by the Shavitt step number\cite{GUGA1} sequences $\{(d_0^{\mu}d_1^{\mu}\cdots)\}|_{\mu=0}^{N_{S,S}^{I-1}}$, which can be arranged in a lexical order
(i.e., $|I\mu\rangle$ precedes $|I\nu\rangle$ if and only if $d_p^{\mu}<d_p^{\nu}$ for $p=\min\{x|d_x^{\mu}\neq d_x^{\nu}\}$), so as to fix the relative ordering $\mu$ uniquely. A 32-bit integer is required to store the relative ordering of CSFs $\{|I\mu\rangle\}$. Since both its BOC and lexical order have to be stored, the memory requirement for storing a CSF is $M_{CFG}+8$ bytes due to memory alignment.
The odering of CSFs is defined as follows: if CSF $|I\mu\rangle$ and $|J\nu\rangle$ stem from the same CFG, then $|I\mu\rangle < |J\nu\rangle$ if $\mu<\nu$;
otherwise, let $Occ_Y$ ($Y=I, J$) be the corresponding occupation array of CFG $|Y\rangle$ and $i = \arg\max_j\{Occ_I[j]\neq Occ_J[j]\}$, then  $|I\mu\rangle < |J\nu\rangle$ if $Occ_I[i] > Occ_J[i]$.
In view of Eq. \eqref{NSS}, a large number of CSFs can be generated from a CFG of high seniority $N_o^I$ but only a few of them may contribute discernibly.
Therefore, it is absolutely necessary to do individual selections of CSFs.

After the above detailed description of TUGA, a gross comparison with the Table-CI approach\cite{TableCI1980,TableCI1986,TableCI1995,TableCIMark2} is in order.
Although both are configuration-driven algorithms for calculating the Hamiltonian matrix elements over randomly selected CSFs, they actually differ significantly.
First of all, Table-CI is closely related to the symmetric group approach (SGA)\cite{SGA1985} in view of its explicit tabulation of line-up permutations,
whereas TUGA is plainly an implementation of the UGA \cite{CSF1,Paldus1980,CSF2},
in terms of the tabulated ROTs rather than the famous graphical representation\cite{GUGA1,GUGA2} (which is efficient
only for well-structured wave functions).
Secondly, TUGA and Table-CI differ in the handling of open-shell orbitals.
While Table-CI always defines CFGs with doubly occupied orbitals sitting together and open-shell orbitals sitting together
(which involves reordering of the orbitals for each CFG), TUGA adopts a universal orbital ordering defined in the very beginning.
The common open-shell orbitals are eliminated in Table-CI by virtue of line-up permutations, leaving
a relatively small number of unique interaction patterns between SDs resulting from non-identical open-shell orbitals. However, the BCCs between CSFs are dependent on the relative positions
of common open-shell orbitals. Therefore, they are not permuted in TUGA but stay where they are after deleting the common zero or doubly occupied orbitals.
This increases merely the lengths of code sequences but not the number of ROTs (NB: every CFG pair corresponds to a ROT).
In essence, the ROTs in TUGA play the same role in classifying unique interacting CFG pairs at the integral level as the Tables defined in Table-CI.
The former are calculated on an as-needed basis, whereas the latter, characterized by
a number of parameters (i.e., $\Delta K$, $P$, $Q$, and $R$ determined by the relative occupations of CFG pairs;
see Ref. \citenum{TableCIMark2} for more details), are created and stored in advance.
Thirdly, TUGA evaluates the Hamiltonian matrix elements directly between CSFs, whereas Table-CI has to assemble them via
transformations of those between SDs (which can be viewed as an effective means to avoid the overly complicated matrix elements of
line-up permutations $Q$ over CSFs, as required by the genuine SGA\cite{SGA1985}).

In short, TUGA has been based on the single, simple fact that the segment values for common zero or doubly occupied orbitals are always one
under the YK phase\cite{Paldus1980}, such that the BCCs are determined solely by those common open-shell orbitals or orbitals with different occupancy numbers (which are
encoded in the code sequences) and can be calculated directly through products of their segment values.

\section{Selection of CSFs}\label{Selection}
The aim of selection is to find a better variational space $P$ by feeding in a guess space $P_0$. This can be achieved in two steps, ranking and pruning.
In the ranking step, proper rank values for the
CSFs in the FOIS $Q$ $(=1-P_0)$ will be evaluated. Those CSFs with rank values larger than
the given ranking-threshold are put into $P_0$, leading to an expanded space $P_1$.
In the pruning step, the Hamiltonian matrix in $P_1$ is first constructed and diagonalized.
Those CSFs with coefficients smaller in absolute value than the given pruning-threshold are then discarded, so as to reduce $P_1$ to $P$.
The procedure is repeated until the energies obtained by diagonalizing
$P$ and $P_0$ are sufficiently close.
Such a two-step selection scheme has been adopted by most sCIs.
Yet, a number of refinements has been introduced in our own implementation\cite{iCIPT2}:
\begin{enumerate}[(1)]
\item\label{Selec1}A combined integral- and coefficient-driven algorithm is used for the selection, viz.
prior to the selection of individual CSFs with their (approximate) first-order coefficients as rank values, a integral-driven screening of unimportant, doubly connected CFG pairs is first carried out
based on upper bounds $\tilde{H}^{IJ}$ of $\langle I\mu|H_2^2|J\nu\rangle$ (which are independent of CSFs, see Table \ref{PaldusType}).
\item\label{Selec2} To minimize the usage of memory,
an increment access of the $Q$ space is invoked, with a dynamically adjusted integral-threshold $\varepsilon_k$ ($=\frac{1}{2}\varepsilon_{k-1})$. For a given $\varepsilon_k$
of iteration $k$ (i.e., $Q[\varepsilon_k]$), the selection is repeated
until $P[\varepsilon_k]$ and $P_0[\varepsilon_k]$ are sufficiently similar in composition (i.e., $\frac{|P_0[\varepsilon_k]\bigcap P[\varepsilon_k]|}{|P_0[\varepsilon_k]\bigcup P[\varepsilon_k]|}\ge 0.95$).
\item\label{Selec3} Natural orbitals (NO) are also
generated dynamically (i.e., for each converged $P[\varepsilon_k]$; for more details, see Algorithm 4 in Ref. \citenum{iCIPT2}).
\item\label{Selec4} A residue-based algorithm\cite{Residue,ASCI2020,ASCI2018PT2} is used for establishing the connections between randomly selected CFGs and CSFs.
\item\label{Selec5} The diagonalization of Hamiltonian matrix is performed with the iVI-Buf(0+3) approach\cite{iVI,iVI-TDDFT},
which has the potential to access directly the roots of a given energy window.
%\item\label{Selec6} Instead of energetic similarity, the composition similarity is used to terminate the selection (i.e., $\frac{|P_0[\varepsilon_k]\bigcap P[\varepsilon_k]|}{|P_0[\varepsilon_k]\bigcup P[\varepsilon_k]|}\ge 0.95$).
\item\label{Selec7} The size of the resulting variational space $P[\varepsilon_k=0]$ is controlled by a single parameter $C_{min}$ (coefficient pruning-threshold), for other parameters have been fixed to conservative values.
\end{enumerate}

It has been demonstrated that the above selection can generate a very compact yet accurate variational space.
However, its efficiency is not yet satisfactory because of the following reasons related to the first four points above:
(a) the adopted ranking criterion is too good (and hence expensive) for the purpose of selection.
(b) Although the incremental access of the $Q$ space is very effective in reducing memory requirement and does not affect the final result, it does result in
additional iterations and hence slow down the selection procedure. (c) The generation of NOs and hence transformation of integrals
are performed too many times, which is unnecessary. (d) The residue-based algorithm is not optimal for identifying the connections between the CFGs in $P$.
In particular, it does not allow for an easy construction of the Hamiltonian matrix in CSR format.
To resolve these issues, we here introduce a simpler ranking criterion, remove the incremental access of $Q$, simplify the generation of NO, and replace
the residue-based algorithm with a particle-hole algorithm for Hamiltonian construction (see Sec. \ref{FastHmat}).

\begin{figure}[!htp]
	\centering
	\includegraphics[width=\textwidth]{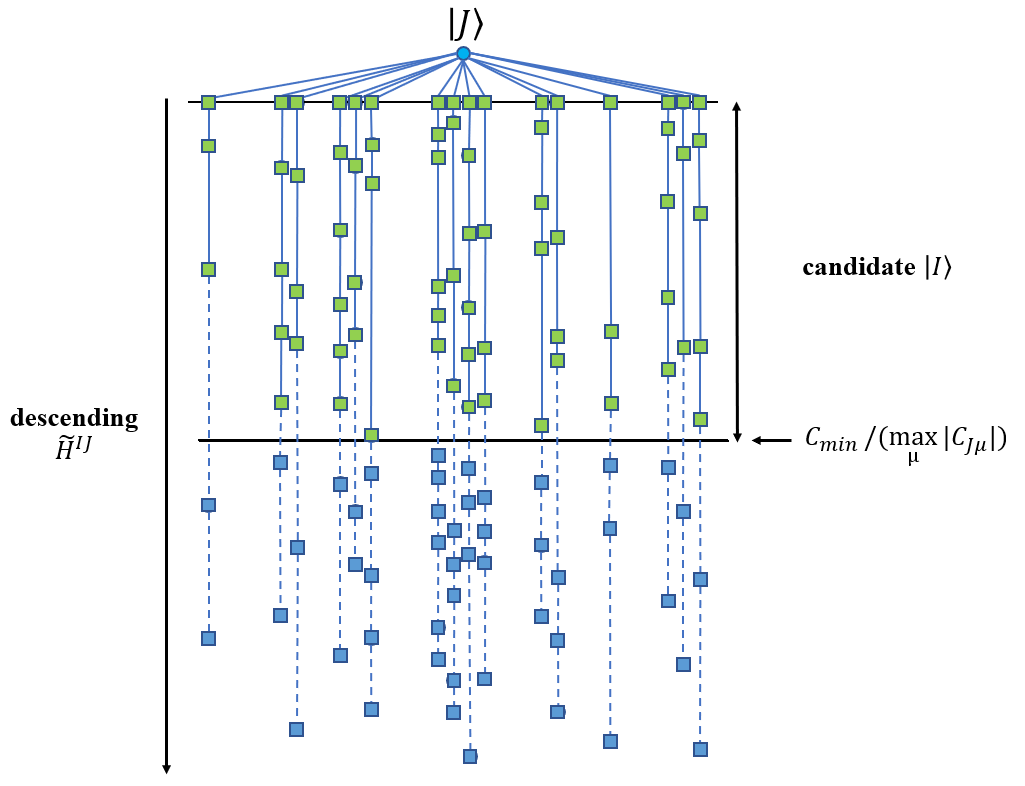}
	\caption{Screening of doubly excited configurations based on upper bounds.}\label{SeleFast}
\end{figure}

In view of many-body perturbation theory (MBPT), the most rigorous ranking of a CSF $|I\mu\rangle \in Q$ is its first-order coefficient in absolute value
\begin{equation}
f(|I\mu\rangle, C_{min}) = \left|\frac{\sum_{|J\nu\rangle\in P_0}H^{IJ}_{\mu\nu}C^J_{\nu}}{E_0-H_{\mu\mu}^{II}}\right|\ge C_{min}, \quad H^{IJ}_{\mu\nu}=\langle I\mu|H|J\nu\rangle,\label{CIPSIrank}
\end{equation}
which is to be called `CIPSI criterion', with CIPSI standing for `configuration interaction with perturbative selection made iteratively'\cite{CIPSIa}.
If those $|J\nu\rangle \in P_0$ with small coefficients are excluded from the sum in the numerator of Eq. \eqref{CIPSIrank}, it may be termed `pruned CIPSI criterion' or simply `ASCI criterion', as adopted by
adaptive sampling CI\cite{ASCI2016}. Noticing that the computational expense of Eq. \eqref{CIPSIrank} results from both the sum in the numerator and the evaluation of the denominator, a dramatically simplified criterion
is considered in heat-bath CI (HBCI)\cite{HBCI2016}, viz.,
\begin{equation}
f(|I{\mu}\rangle,|J\rangle, \epsilon_Q)=\max_{\nu}|H^{IJ}_{\mu\nu}C^{J}_{\nu}|\stackrel{\mathrm{CSF}\rightarrow\mathrm{SD}}\longrightarrow \max_{\nu} |H_{\mu\nu}C_{\nu}|
\ge\epsilon_{Q},\label{HBCIrank}
\end{equation}
in the basis of SDs though. This way, those determinants $\{|\mu\rangle\}$ that are doubly excited from $|\nu\rangle\in P_0$ are never accessed if $|H_{\mu\nu}|<\epsilon_Q/|C_{\nu}|$\cite{HBCASSCF}.
However, such integral-driven selection leads usually to a variational space that is much less compact than that by the coefficient-driven selection \eqref{CIPSIrank}, particularly when a large basis set is used.
A possible remedy to this is to introduce\cite{HBCI-Cr2} an approximate denominator to condition \eqref{HBCIrank}, without sacrificing the efficiency.
Our previous `iCI criterion'\cite{iCIPT2} amounts
to combining \emph{dynamically} these integral- and coefficient-driven algorithms for selecting doubly excited CSFs, viz.,
\begin{equation}
f(|I\mu\rangle, \varepsilon_k, C_{min})= \left|\frac{\sum_{J\in P_0^{(k)}}^{(\varepsilon_k)}
(\sum_{\nu,|J\nu\rangle\in P_0^{(k)}}H^{IJ}_{\mu\nu}C^{J(k)}_{\nu})}{E^{(k)}-H_{\mu\mu}^{II}}\right|\ge C_{min}, \label{iCIrank}
\end{equation}
where $\varepsilon_k$ is a dynamically adjusted integral-threshold and the summation over CFG $|J\rangle\in P_0^{(k)}$ is subject to the following condition
\begin{equation}
\max_{\nu}|\tilde{H}^{IJ}C_{\nu}^{J(k)}|\ge\varepsilon_k=\frac{1}{2}\varepsilon_{k-1},\quad |J\rangle\in P_0^{(k)}.\label{iCI-J}
\end{equation}
The original motivation for this particular criterion is to reduce memory usage as much as possible and meanwhile to end up with a variational space that is as compact as possible.
It appears that both are not really necessary: the memory requirement in the PT2 step is much larger and a slight sacrifice of the compactness can be overcompensated by the dramatic gain in efficiency.
Therefore, we introduce here a new `iCI criterion', that is,
for a given $|I\mu\rangle \in Q=1-P_0$,
if there exists a $|J\rangle\in P_0$ for which $f(|I\mu\rangle,|J\rangle,C_{min})$ is true,
then $|I\mu\rangle$ is selected. The boolean function $f(|I\mu\rangle,|J\rangle,C_{min})$ is defined as
\begin{enumerate}[(A)]
	\item\label{Case1} If $|I\rangle$ is identical with or singly excited from $|J\rangle$, then
	\begin{equation}
	f(|I\mu\rangle,|J\rangle,C_{min}) = \left(\max_{\nu}(|H^{IJ}_{\mu\nu}C_{J\nu}|)\geq C_{min}\right)\text{ and } \left(\max_{\nu}\left(\left|\frac{H^{IJ}_{\mu\nu}C_{J\nu}}{E_{0}-H^{II}_{\mu\mu}}\right|\right)\geq C_{min}\right); \label{iCI_Rank_1}
	\end{equation}
	\item\label{Case2} If $|I\rangle$ is doubly excited from $|J\rangle$, then
	\begin{eqnarray}
	f(|I\mu\rangle,|J\rangle,C_{min}) &=&\left(\max_{\nu}(|\tilde{H}^{IJ}C_{J\nu}|)\geq C_{min}\right)\text{ and } \left(\max_{\nu}(|H^{IJ}_{\mu\nu}C_{J\nu}|)\geq C_{min}\right)\nonumber\\
& &\text{ and }  \left(\max_{\nu}\left(\left|\frac{H^{IJ}_{\mu\nu}C_{J\nu}}{E_{0}-H^{II}_{\mu\mu}}\right|\right)\geq C_{min}\right). \label{iCI_Rank_2}
	\end{eqnarray}
\end{enumerate}
Literally, for case (\ref{Case1}), loop over all $|I\mu\rangle$ in $Q$ and evaluate $H^{IJ}_{\mu\nu}$ for all CSFs $|J\nu\rangle\in P_0$.
If $\max_{\nu}(|H^{IJ}_{\mu\nu}C_{J\nu}|)$ is larger than $C_{min}$ then evaluate $H^{II}_{\mu\mu}$; otherwise discard $|I\mu\rangle$. If $\max_{\nu}(|\frac{H^{IJ}_{\mu\nu}C_{J\nu}}{E_0-H^{II}_{\mu\mu}}|)$ is larger than $C_{min}$ then $|I\mu\rangle$ is selected.
As for case (\ref{Case2}), only those doubly excited CFGs $|I\rangle$ with $\tilde{H}^{IJ}$ larger than $C_{min}/\max_{\nu} |C_{J\nu}|$ need to be generated
(i.e., those unimportant ones are never touched, as illustrated in Fig. \ref{SeleFast}). For such $\{|I\rangle\}$, the remaining step is the same as case (\ref{Case1}).

It should be clear that, for the same $C_{min}$,
the revised `iCI criterion' \eqref{iCI_Rank_1}/\eqref{iCI_Rank_2} will yield a smaller number of CSFs than the original one \eqref{iCIrank}/\eqref{iCI-J}, for the latter may
accumulate those CSFs of small coefficients until the condition is fulfilled (cf. the second sum in Eq. \eqref{iCIrank}), whereas the former just picks up the CSF with the largest coefficient.
Note in passing that, to be compared directly with this `iCI criterion', the `HBCI criterion' \eqref{HBCIrank} should be rewritten as
\begin{enumerate}[(a)]
	\item If $|I\rangle$ is identical with or singly excited from $|J\rangle$, then
	\begin{equation}
	f(|I\mu\rangle,|J\rangle,C_{min}) = \max_{\nu}(|H^{IJ}_{\mu\nu}C_{J\nu}|)\geq C_{min};	\label{HBCI_Rank_1}
	\end{equation}
	\item  If $|I\rangle$ is doubly excited from $|J\rangle$, then
	\begin{equation}
	f(|I\mu\rangle,|J\rangle,C_{min}) =\left(\max_{\nu}(|\tilde{H}^{IJ}C_{J\nu}|)\geq C_{min}\right)\text{ and } \left(\max_{\nu}(|H^{IJ}_{\mu\nu}C_{J\nu}|)\geq C_{min}\right),\label{HBCI_Rank_2}
	\end{equation}
\end{enumerate}
which amounts to skipping the final step in Eqs. \eqref{iCI_Rank_1}/\eqref{iCI_Rank_2}.

\begin{figure}[!htp]
	\centering
	\includegraphics[width=\textwidth]{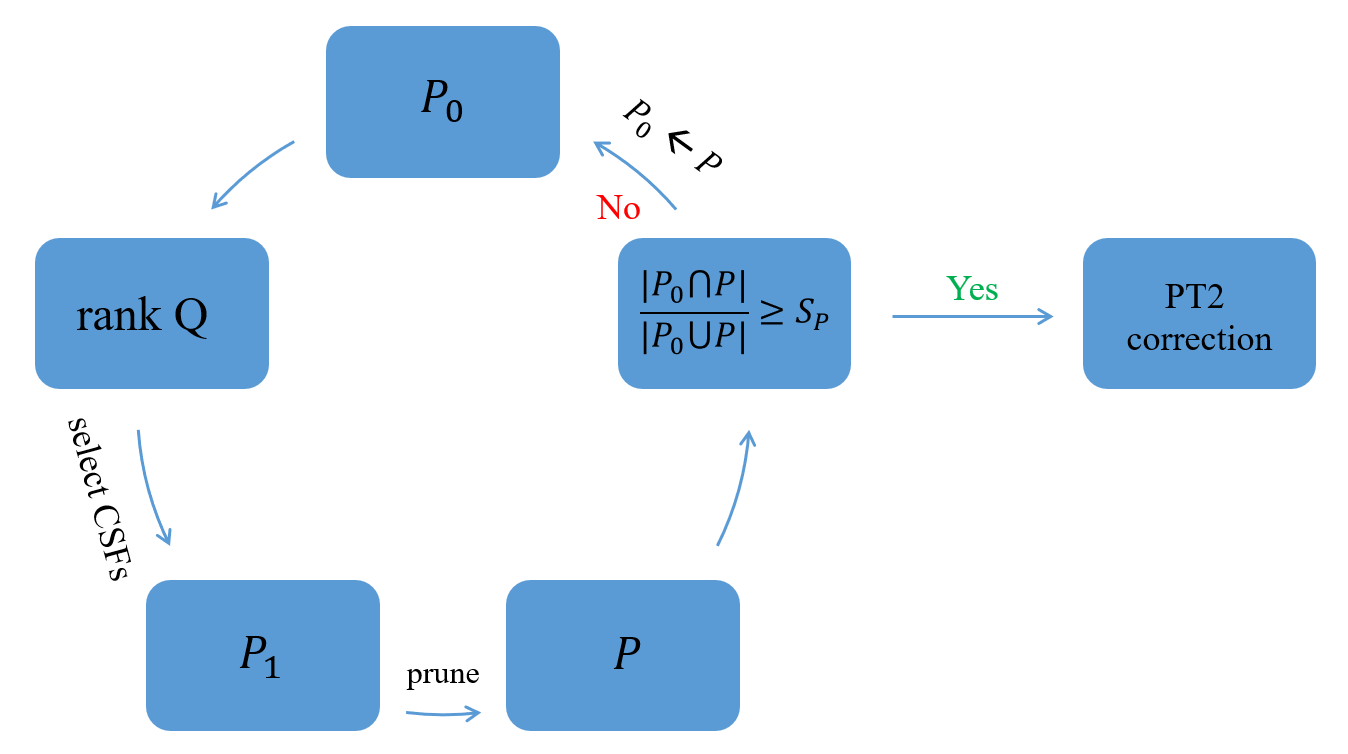}
	\caption{Flowchart for iCIPT2.}\label{SeleWorkFlow}
\end{figure}

The overall selection procedure is illustrated in Fig. \ref{SeleWorkFlow}. Four additional points still deserve to be mentioned:
\begin{enumerate}[(1)]
\item For multiple states, the selection will be carried out in a state-collective way. That is, the boolean function $f_k$ defined in Eq. \eqref{CIPSIrank},
\eqref{iCI_Rank_1}/\eqref{iCI_Rank_2} or \eqref{HBCI_Rank_1}/\eqref{HBCI_Rank_2}
is examined for each state $k$ and $|I\mu\rangle$ is selected if $\exists k$ and $|J\nu\rangle$ s.t. $f_k=\mathrm{true}$.
\item Duplicates in the selected CSFs $\{|I\mu\rangle\}\in Q$ are removed as follows.
The for-loop over CFGs $|J\rangle$ in $P_0$ is distributed among working threads. For each thread, once the memory usage has reached a given size (e.g., 2 GB),
the selected CSFs $\{|I\mu\rangle\}$ are sorted immediately,
 with duplicates removed by a single loop. To expedite this, the
CSFs are classified according to their seniorities and highest occupied orbitals.
The surviving CSFs $\{|I\mu\rangle\}$ are then compared with those in $P_0$ to further remove the duplicates.
Once the loop over $|J\rangle$ is done, the surviving CSFs $\{|I\mu\rangle\}$ from each thread
are merged to the main thread and a final removal of duplicates is then executed.
\item The selection is terminated once the measure of composition similarity, $\frac{|P_0\bigcap P|}{|P_0\bigcup P|}$, exceeds the threshold $S_P$ (e.g., 0.95).
\item If desired, NOs can be generated in a partially dynamic manner,
    e.g., going from $C^\prime_{min}=4C_{min}$ to $2C_{min}$ and finally to $C_{min}$.
    For each $C_{min}'$, the above selection is performed to find the corresponding variation space $P$, from which the NOs can be generated.
   After integral transformation, the Hamiltonian matrix is reconstructed and diagonalized, with unimportant CSFs pruned. This extra step is important for the expansion coefficients of the wave function are affected by orbital rotations. The NOs generated this way usually give rise to a more compact variational space than those
   generated directly at the target $C_{min}$.
\end{enumerate}

% The original NO generation:
%In the case of FCI calculation, the quality of single particle orbitals has no effect on the final wavefunction. In an SCI calculation, including iCIPT2, the quality of single particle orbitals have great impact on the convergence of a simulation with respect to $C_{min}$. Hence single particle orbitals can be optimized to improve the convergence. In the original iCIPT2, after each middle iteration, it constructs natural orbitals by diagonalizing 1-RDM then determines whether to update $\varepsilon_{cps}$. It stops natural orbital rotation if single particle orbitals are thought to be converged. The condition for the convergence of NOs is as follows. At $k$--th middle iteration, the NOs $\{\phi_p^{(k+1)}\}$ and $\{\phi_p^{(k)}\}$ are related by $\phi_p^{(k+1)}=\sum_{q}c_{qp}\phi_q^{(k)}$. If $\exists q \text{ s.t. } |C_{qp}|\geq 0.99$, $\phi_p^{(k+1)}$ is considered to be converged (the same as $\phi_q^{(k)}$). If the sum of the occupation numbers of the converged NOs deviates from the total number of correlated electrons by less than 0.05, $\{\phi_p^{(k+1)}\}$ are considered to be the same as $\{\phi_p^{(k)}\}$.  In the original iCIPT2, the wavefunction grows larger when $\varepsilon_{cps}$ approaches zero (incremental access of $Q$) while the single particle orbitals are optimized towards the optimal natural orbitals.

\section{Connections between selected CSFs}\label{FastHmat}
Having identified the most important CSFs spanning space $P$, their connections must be established for constructing and updating the Hamiltonian matrix.
In the previous implementation of iCIPT2\cite{iCIPT2}, we adopted the residue array-based algorithm\cite{Residue,ASCI2020,ASCI2018PT2}.
The $n$th-order residues of an $N_{e}$-electron CFG $|I\rangle$ are those $(N_{e}-n)$-electron CFGs that can be generated by removing $n$ electrons from $|I\rangle$ in all possible ways.
A nice point of residues lies in that they provide connection information on the parent CFGs. For instance,
different CFG pairs sharing the same first-order residue (FOR) are identical or singly connected, while those sharing the same second-order residue (SOR) are identical, singly or doubly connected.
However, for the purpose of constructing and updating the Hamiltonian matrix in the $P$ space, the residues have some drawbacks:
(1) the number of SORs scale linearly with respect to the number $N_{cfg}$ of CFGs in $P$ but scale quadratically with respect to the number $N_{e}$ of
correlated electrons, such that both the consumption of memory and the repeated update of the SORs (which is necessary during
the iterative selection) are unfavorable when $N_{e}$ is large. (2) There may exist numerous redundant residues that do not provide any
connection information, especially in the presence of many doubly occupied orbitals. (3)
More seriously, the CFGs connected to a given CFG may be scattered widely in location,
such that it is not easy to construct the Hamiltonian matrix in CSR format for efficient matrix-vector multiplications (unless a sorting step is taken but which itself is not cheap).
It turns out that all these issues can be resolved by going to a particle-hole representation of the CFGs in $P$. The holes and particles are
just the doubly occupied and the remaining singly or zero occupied orbitals in the common reference CFG $|R\rangle$ (which is already adopted in Eq. \eqref{H-diag-3}), respectively.
Then, the array OrbOccBinary representing a CFG can be decomposed into a hole (HoleStr) and a particle part (PartStr), such that connections between CFG pairs
can be identified from those between HoleStr's and between PartStr's. To this end,
we first define single transitions within the hole or particle orbitals as single inner excitations (SIE), whereas those between the two sets as single outer excitations (SOE) or de-excitations (SDE).
While SIEs do not change the lengths of HoleStr and PartStr, a SOE (SDE) will increase (decrease) their lengths by one.
Two HoleStr's (PartStr's) are connected only if their lengths differ less than three.
A given HoleStr/PartStr StrA  with length \textit{len} can be connected with a HoleStr/PartStr StrB in eight possible ways,
\begin{enumerate}[1.]
	\item StrB with length \textit{len}+2 and two SOEs;
	\item StrB with length \textit{len}+1 and one SOE;
	\item StrB with length \textit{len}+1, one SIE and  SOE;
	\item StrB with length \textit{len} and one SIE;
	\item StrB with length \textit{len} and two SIE;
	\item StrB with length \textit{len}-1 and one SDE;
	\item StrB with length \textit{len}-1, one SIE and one SDE;
	\item StrB with length \textit{len}-2 and two SDEs,
\end{enumerate}
in addition to the case (denoted as 0) when StrB is identical with StrA.

The data structure of HoleStr is very simple. It consists of three parts: (1) an array of 64-bit integers recording the corresponding HoleStr;
(2) eight hole connection arrays $\mathcal{C}_{ph}$ recording the above connections to other HoleStr's; (3) one CFG array $\mathcal{G}_{ph}$ recording the indices of CFGs sharing the same HoleStr.
The data structure of PartStr is exactly the same. The connections between CFG pairs can then be identified in a simple way, since
singly connected CFG pairs can only be one of the three single-connection types (SCT)
$\{(2,2), (4,0), (6,6)\}$, whereas doubly connected CFG pairs can only
be one of the eight double-connection types (DCT) $\{(1,1), (2,3), (3,2), (4,4), (5,0), (6,7), (7,6), (8,8)\}$.
For instance, for a given CFG $|I\rangle$ with HoleStr HoleI and PartStr PartI, one can first find all HoleJ's and PartJ's
according to the SCTs.
Then, for each pair of HoleJ and PartJ, the intersection of their $\mathcal{G}_{ph}$'s contains all the CFGs that are singly connected to CFG $|I\rangle$.
The doubly connected CFG pairs can be identified in the same way.

The Hamiltonian matrix can be constructed as follows: (A) loop over all CFGs $|I\rangle\in P$. (B) For each $|I\rangle$, find all CFGs $|J\rangle\in P$ connected with $|I\rangle$
but with indices smaller than that of $|I\rangle$. (C) Loop over $|J\rangle$ and calculate the Hamiltonian
matrix elements between $|I\mu\rangle$ and $|J\nu\rangle$, which can readily be compressed in CSR format because the indices are now contiguous.
As for the update of the Hamiltonian matrix, suppose that the old and new CSF spaces are $P_1\subset P_2$ and $P_2$, respectively, and the
CSFs in $P_2-P_1$ have been grouped according to their CFGs. After updating the connections between HoleStr's and between PartStr's,
just loop over all newly added CFGs $|I\rangle$ in $P_2-P_1$ and do (B) and (C) above.

In summary, as far as the handling of the $P$ space is concerned, the present particle-hole algorithm is advantageous
over the residue-based algorithm\cite{Residue,ASCI2020,ASCI2018PT2} in several aspects, especially when $N_e$ is large. First of all,
the metadata of the HoleStr's and PartStr's is much smaller in size than that of the FORs and SORs, which is reflected by
the much reduced memory consumption.
For instance, for a variational space with 1164710 unique CFGs or 3059395 CSFs in the all-electron calculation of \ce{Cr2} with the Ahlrichs SV basis set\cite{AhlrichsSV},
the memory cost of the HoleStr's and PartStr's is only 0.55 GB,
whereas those of the FORs and SORs are 2.48 GB and 29.1 GB, respectively.
%Because of this, the update of HoleStr's and PartStr's is also much easier.
%The cost for updating the indices of CFGs stored in HolePtr's and PartStr's is $O(N_{cfg})$ since each CFG corresponds exactly to one HoleStr and one PartStr.
%In contrast, the costs for updating the first- and second-order residues are
%$O(N_{cfg}*N_{elec})$ and $O(N_{cfg}*N_{elec}^2)$, respectively.
Secondly, the simple structure of the HoleStr's and PartStr's allows for an easy construction of
the Hamiltonian matrix in CSR format. For the case of \ce{Cr2}, the Hamiltonian construction is speeded up by $2.5\times$.
Thirdly, the connections between HoleStr's and between PartStr's can readily be reutilized.
For example, when updating the Hamiltonian matrix from $P_1$ to $P_2$ mentioned previously, if
some CSFs of a CFG in $P_2-P_1$ are present in $P_1$, the connections of this CFG to other CFGs in $P_1$ have already been encoded
in the connections between HoleStrs and between PartStrs of $P_1$.
More interestingly, such connections can also
be shared by CSF spaces of different spatial and/or spin symmetries, which facilitates the simultaneous calculation of several
states of different spatial and/or spin symmetries with a common set of orthonormal orbitals.

As a final note, it deserves to be mentioned that, because of the random nature of the $P$ space, the number of CFG pairs therein sharing the same Rot\_Code is usually very small,
such that the reutilization of BCCs is ineffective: it cannot compensate the overhead necessary for sorting the CFG pairs (which requires both substantial memory and synchronous operations).
Instead, recalculating the BCCs whenever needed turns out to be more efficient. However, the situation is different for the PT2 step, where
the reutilization of BCCs by TUGA is essential.

%\subsection{Other Problems}
%
%Meanwhile, in the original iCIPT2, there are some problems irrelevant to residues: (1) In the original iCIPT2, the new space $P_1$ and the old space $P_0$ are not forced to satisfy the condition $P_0\subset P_1$. To update Hamiltonian matrix, some matrix elements in $H_0$ have to be removed with the row/column index being updated.
%%which is time-consuming and unnecessary.
%In the fast iCIPT2, $P_0$ and $P_1$ are forced to satisfy $P_0\subset P_1$ and the new added CSFs are simply appended to $P_0$. (2) In order to reutilize coupling coefficients, all calculation records $\{I,J,ROT\_Code,integrals\}$ have to be generated before calculating matrix elements, which not only leads to a memory comsumption peak but also reduces the parallel efficiency greatly. In the fast iCIPT2, once the connected CFG pair is identified, the corresponding matrix elements are calculated. Our numerical tests show that this does not reduce the efficiency of constructing Hamiltonian matrix but enhance the parallel efficiency.

\section{Constraint-Based ENPT2} \label{PT2}
%In common with most other SCI+PT method, the perturbative correction is computed using Epstein Nesbet perturbation theory. The variational wavefunction is used to define the zero-order Hamiltonian, $H_0$ and the perturbation $V$ as
%
%\begin{equation}
%\begin{split}
%	H_0 &= \sum_{|I\mu\rangle,|J\nu\rangle\in P} H^{IJ}_{\mu\nu}|I\mu\rangle\langle J\nu|+\sum_{|I\mu\rangle\notin P}H^{II}_{\mu\mu}|I\mu\rangle\langle I\mu|\\
%	V   &= H-H_0
%\end{split}
%\end{equation}
As long as the selectively determined variational space $P$ and hence the zeroth-order eigenpairs $\{E_k^{(0)},|\Psi_k^{(0)}\rangle\}$ are good enough, the remaining dynamic correlation can be accounted for accurately at the lowest level of theory.
Here, we adopt the state-specific ENPT2
\begin{eqnarray}
E_{c,k}^{(2)} &=& \sum_{|I\mu\rangle\in Q}\frac{|\langle I\mu|H|\Psi^{(0)}_k\rangle|^2}{E^{(0)}_k-H^{II}_{\mu\mu}}\label{ENPT2Defa}\\
&=&\sum_{|I\mu\rangle\in Q} \frac{\left|\sum_{|J\nu\rangle\in P}H^{IJ}_{\mu\nu}C^J_{\nu,k}\right|^2}{E_k^{(0)}-H_{\mu\mu}^{II}}.\label{ENPT2Defb}
\end{eqnarray}
Before evaluating this energy expression, several important points should be observed. Firstly, what the interaction really does is to excite
 one or two electrons from space $P$ to space $Q$ in all possible ways, leaving $N_e-1$ or $N_e-2$
CFGs in $P$, which are nothing but the respective FORs and SORs mentioned in the previous section. Reversely, the excited CFGs can be
generated by inserting one and two electrons into the FORs and SORs, respectively.
Therefore, the FORs and SORs provide natural connections
between the two spaces and are hence the proper organizing units. Secondly, the number of CFG pairs sharing the same Rot\_Code can be very large, such that
is essential to reutilize the BCCs by TUGA.
Thirdly, the summation over $|I\mu\rangle\in Q$ can be extended to include also the CSFs in $P$, by precomputing and finally subtracting their
PT2-like energy $\bar{E}_{c,k}^{(2)}$\cite{ASCI2018PT2}, viz.,
\begin{align}
E_{c,k}^{(2)} &=\tilde{E}_{c,k}^{(2)}-\bar{E}_{c,k}^{(2)},\label{ENPT2diff}\\
\tilde{E}_{c,k}^{(2)}&= \sum_{|I\mu\rangle\in W} \frac{\left|\sum_{|J\nu\rangle\in P, |J\nu\rangle\ne |I\mu\rangle}H_{\mu\nu}^{IJ}C_{\nu,k}^J\right|^2}{E_k^{(0)}-H_{\mu\mu}^{II}},\quad W=P\cup Q,\label{ENPT2Qt}\\
\bar{E}_{c,k}^{(2)}&=\sum_{|I\mu\rangle\in P} \frac{\left|\langle I\mu|H|\Psi_k^{(0)}\rangle-C_{\mu,k}^I H^{II}_{\mu\mu}\right|^2}{E_k^{(0)}-H_{\mu\mu}^{II}}\label{ENPT2Pa}\\
                   &=\sum_{|I\mu\rangle\in P} (C_{\mu,i}^I)^2 (E_i^{(0)}-H^{II}_{\mu\mu}),\label{ENPT2P}
\end{align}
where use of the relation $\langle I\mu|H|\Psi_k^{(0)}\rangle=C_{\mu,k}^I E_k^{(0)}$ has been made when going from Eq. \eqref{ENPT2Pa} to Eq. \eqref{ENPT2P}.
The negative term in the numerator of Eq. \eqref{ENPT2Pa} arises from the fact that the diagonal terms have been excluded in Eq. \eqref{ENPT2Qt}. This way,
there is no need to double check whether the excited CSFs belong to $Q$ or $P$, which is every expensive.
Fourthly, since there is no coupling between the $|I\mu\rangle$ functions, the whole $W$ space
can be decomposed into a series of disjoint subspaces $\{W_i\}_i^{N_s}$, so as to reduce memory requirement and meanwhile facilitate the parallelization.

Following the idea of constraint PT2\cite{ASCI2018PT2}, a CFG subspace $W_i$ of $W$ can be defined by a constraint consisting of $L_c$ highest occupied orbitals
(say, $p_1, p_2, \cdots, p_{L_c}$ in ascending order) as well as their occupation numbers (say, $n_{p_1},n_{p_2},\cdots,n_{p_{L_c}}$).
The number $N_s$ of such subspaces is bounded by $C_{N_{orb}}^{L_c}\times 2^{L_c}$. To achieve this, the FORs and SORs of $P$ are first generated and sorted, with the unique ones recorded in arrays $\mathcal{R}_{1}$ and $\mathcal{R}_{2}$, respectively.
Here, each unique residue is associated with an array of records \textit{\{indx,orbj,orbl\}} to trace how it arises: \textit{indx} records the CFG in $P$, whereas \textit{orbj} and  \textit{orbl} record
the orbitals from which electrons are removed (i.e., the $j$ and $l$ indices of $e_{ij,kl}$).
Note that \textit{orbj} can be equal to \textit{orbl} for a SOR, whereas only \textit{orbsj} is needed for a FOR. Then, those FORs and SORs that
can generate the CFGs belonging to $W_i$ must be identified.
Since the occupation pattern of a CFG in $W_i$ must be $(0,\cdots,0,n_{p_{L_c}},0,\cdots,0,n_{p_{L_c-1}},0,\cdots,0,n_{p_1})$ for orbitals of indices not smaller than $p_1$,
a valid residue with occupation numbers $\{m_n\}_{n=0}^{N_{orb-1}}$ must be subject to the restrictions
 $m_p\leq n_p$ for $p\geq p_1$ and $R_t=\sum_{p=p_1}^{N_{orb}-1}(n_p-m_p)\in [0,2]$.
For the case of FORs,
$R_t$ can be either 0 or 1, which are labeled c1 and c2, respectively. For the former, $n_p=m_p$ for all $p\geq p_1$. For the latter,
there exists one orbital $q$ such that $n_{q}=m_{q}+1$ for $q\in[p_1,p_{L_c}]$ and  $n_p=m_p$ for $p\geq p_1$ other than $q$.
For the case of SORs, $R_t$ can also be 2, which leads to two additional conditions:
(a) c3: if there exists $q\geq p_1$ with $n_q=2$ and $m_q=0$, then $n_p=m_p$ for $p\geq p_1$ other than $q$.
(b) c4: if there exist $q$ and $r$ with $n_q-m_q=1$ and $n_r-m_r=1$, then $n_p=m_p$ for $p\geq p_1$ other than $q$ and $r$.
The valid FORs and SORs can be classified according to their occupation patterns for orbitals $p\geq p_1$, such that those of
the same occupation pattern are located contiguously in $\mathcal{R}_1$/$\mathcal{R}_2$, thereby forming a segment.
The numbers of such patterns/segments are
$1+C_{L_c}^1$ for FORs (1 for c1 and $C_{L_c}^1$ for c2) and $1+C_{L_c}^1+C_{L_c}^1+C_{L_c}^2$ for SORs
($C_{L_c}^1$ for c3 and $C_{L_c}^2$ for c4).
The head $R_{min}$ (tail $R_{max}$) of a segment is determined by distributing the remaining $N_e-\sum_{p\ge p_1}^{N_{orb}-1} m_p-N_R$ electrons ($N_R=1$ for FORs and 2 for SORs)
to orbitals as close to (far away from) $p_1$ as possible. They can be used to locate the relevant residues by binary search of the segments.
For a given residue, each way of adding electrons will generate a CFG $|I\rangle\in W_i$, whose connections with the CFGs $\{|J\rangle\}$ in $P$ can
be established by looping over the records \textit{\{indx,orbj,orbl\}} associated with the residue (NB: each \textit{indx}
corresponds to a CFG $|J\rangle$). Such pair connection records (PCR) are stored in array $\mathcal{C}_{IJ}$, which is to be sorted to group those records sharing the same $|I\rangle$ together.
Note in passing that in principle only the SORs are needed to generate such PCRs. However, very many duplicate singly connected CFG pairs can be generated from the SORs and
the removal of them can be very costly. Instead, it is more favorable to take the (non-duplicate) singly connected CFG pairs from the FORs and simply dump the ones from the SORs.
The price to pay is just some extra memory. For a given PCR, an interaction record $\{ROT\_Code, I, J, integrals\}$
is further needed to calculate $\sum_{\nu}\langle I\mu|H|J\nu\rangle C_{\nu,k}^J$ for all CSFs of $|I\rangle\in W_i$.
It can be constructed by first identifying the ROT for $(I,J)$ and then fetching the integrals based on ROT\_Orb.
All the interaction records are stored in array $\mathcal{V}_{IJ}$, which is to be sorted according to ROT\_Code
to reuse the BCCs by TUGA.

At this moment two key factors should be observed:
(1) the sizes of different subspaces $W_i$ may differ by orders of magnitude, which renders the load in parallelization extremely imbalanced on one hand and the reutilization of BCCs of very small subspaces virtually impossible.
(2) Both the generation and sorting of the connection and interac-
tion data can be very expensive. The former can be resolved by precomputing the sizes of subspaces, so as to merge small subspaces together or further split very large subspaces into smaller ones. Although not very cheap, the expense of this null step is overcompensated by the gain in parallelization efficiency. As for the latter, the red-black tree-based algorithm adopted in our previous implementation\cite{iCIPT2} turns out to be inefficient: too much time is wasted on the searching and insertion, due to the fact that the nodes in red-black trees are scattered in memory, thereby resulting in a very low cache-hit rate. Since array appending and sorting are more efficient than tree insertion and searching, we here employ arrays instead of trees for the data. What is essential is then how to impose a particular structure to the array of raw data, according to which a possibly optimal sorting algorithm can be found. For instance, if the array occurs automatically as sorted subarrays with an equal length, the Timsort algorithm\cite{Timsort,TimsortCode} would be the best choice (NB: Timsort appears to work well even if some subarrays have different lengths).
To this end, the above connection array $\mathcal{C}_{IJ}$ is to be generated as follows:
(1) for the c2 segment of FORs or the c3 and c4 segments of SORs, the electron-accepting orbitals are already fixed and are the same for all the residues within the segment.
Since the FORs and SORs are already sorted, the PCRs generated from such segment are automatically sorted.
(2) For the c1 segment of FORs or the c2 segment of SORs, one electron should be added to orbital $s\in[0,p_1-1]$. If the loop over $s$ proceeds in descending order, the
generated CFGs will be in ascending order, leading to ordered PCRs for each residue. Therefore, the PCRs for such segment
consist of several ordered subarrays, thereby well suited for the Timsort algorithm.
(3) For the c1 segment of SORs, two electrons should be added to orbitals $i, k\in[0,p_1-1]$ with $k\geq i$ (NB: $i,k$ are
the indices of $e_{ij,kl}$). If the outer loop over $k$ and the inner loop over $i$ for each $k$ both proceed in descending order,
the PCRs will be of the same structure as case (2) and hence fit the Timsort algorithm as well.
After looping over all residue segments, the whole $\mathcal{C}_{IJ}$ array will consist of several sorted subarrays and can finally be sorted by Timsort.
In contrast, the interaction array $\mathcal{V}_{IJ}$ cannot be prepared in any particular structure and is hence sorted simply by std::sort() in C++.

The efficacy of the above PT2 algorithm can quickly be revealed by again taking \ce{Cr2}/SV as an example.
It turns out that the smaller the $C_{min}$ (i.e., the larger the variational space $P$), the larger
the speedup of PT2 over the previous, red-black tree-based implementation\cite{iCIPT2}, amounting to 21$\times$ for $P$ with 1.1$M$ CSFs.
If desired, two additional cutoffs can be employed for the PT2 correction:
(1) those doubly excited CFGs $|I\rangle$ with $\tilde{H}^{IJ} \max_\nu |C^J_{\nu,k}|$ smaller than a threshold (e.g., $10^{-3}C_{min}$) can be neglected.
(2) Those CSFs $|I\mu\rangle$ with $|\langle I\mu|H|\Psi_k^{(0)}\rangle|$ smaller than a threshold (e.g., $5\times10^{-9}$) can further be neglected.

\section{Results and discussion}\label{Results}
All the calculations were performed with the BDF program package\cite{BDF1,BDF2,BDF3,BDFECC,BDFrev2020} on one node with 4 Intel(R) Xeon(R) Gold 6240 CPUs (in total 72 physical cores) and 768 GB memory.
\subsection{Comparison of different ranking criteria}
\begin{figure}[!htp]
\centering
\begin{tabular}{cc}
\includegraphics[width=0.5\textwidth]{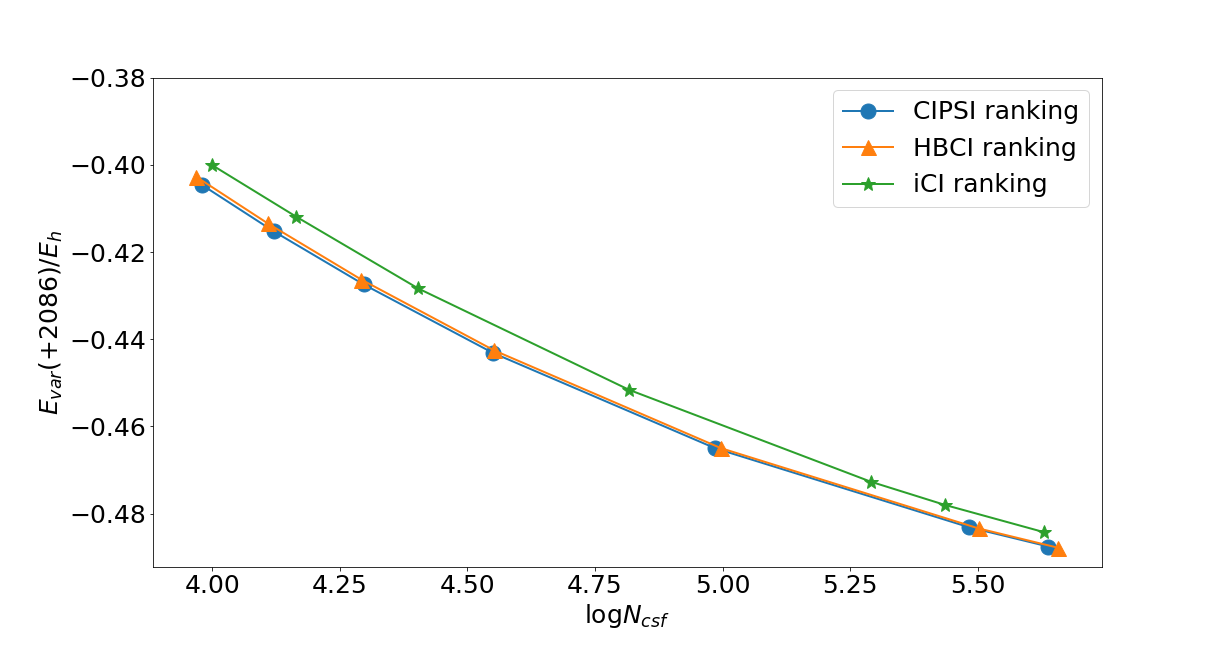}&\includegraphics[width=0.5\textwidth]{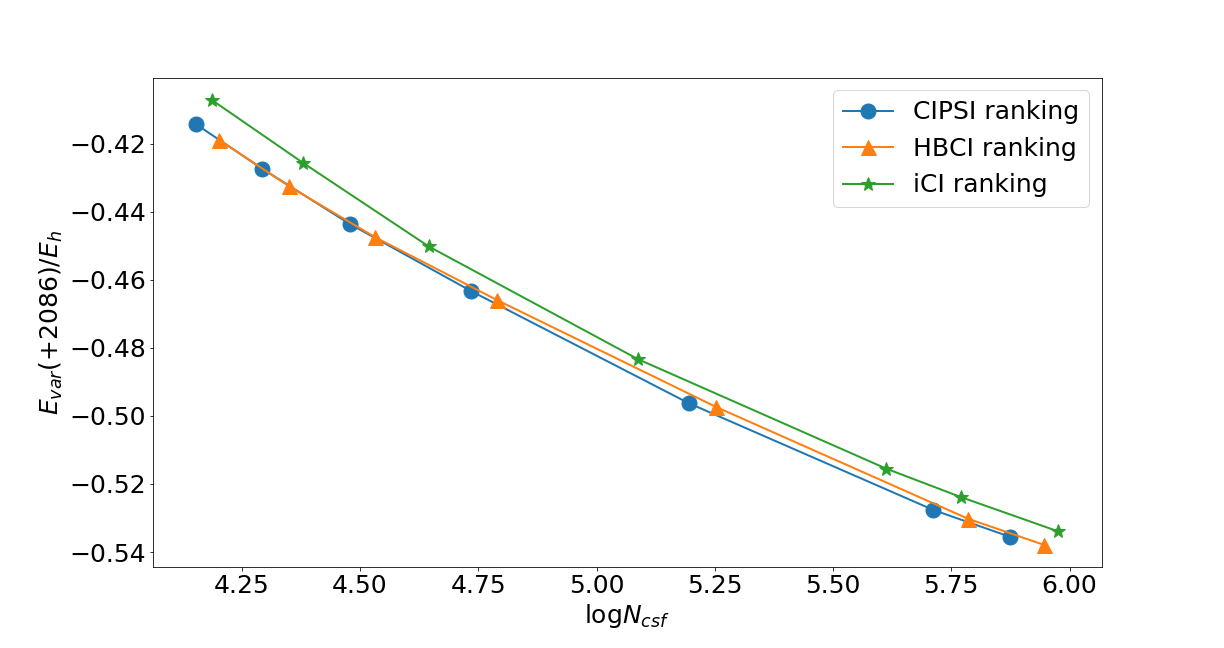}\\
(a) $E_{var}$ at $R_{eq}$ &(b) $E_{var}$ at $2.0R_{eq}$\\
\end{tabular}
\caption{The all-electron variational energy $E_{var}$ of \ce{Cr2}/SV as function of the number $N_{csf}$ of CSFs selected by the CIPSI \eqref{CIPSIrank},
iCI \eqref{iCI_Rank_1}/\eqref{iCI_Rank_2} and HBCI \eqref{HBCI_Rank_1}/\eqref{HBCI_Rank_2} ranking criteria.}
\label{CR2}
\end{figure}

\begin{figure}[!htp]
	\centering
	\includegraphics[width=\textwidth]{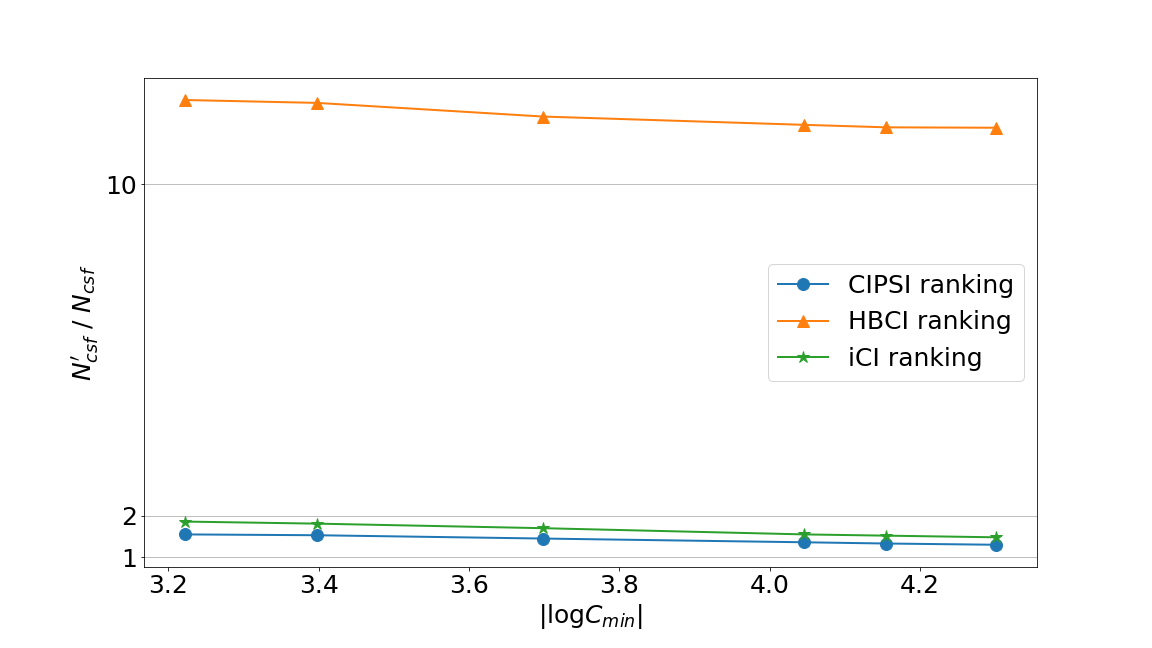}
	\caption{The size ratio $N_{csf}^\prime/N_{csf}$ between the unpruned and pruned variational spaces determined by the CIPSI \eqref{CIPSIrank},
iCI \eqref{iCI_Rank_1}/\eqref{iCI_Rank_2} and HBCI \eqref{HBCI_Rank_1}/\eqref{HBCI_Rank_2} ranking criteria. }
	\label{SizeRatio}
\end{figure}

\begin{figure}[!htp]
	\centering
	\includegraphics[width=\textwidth]{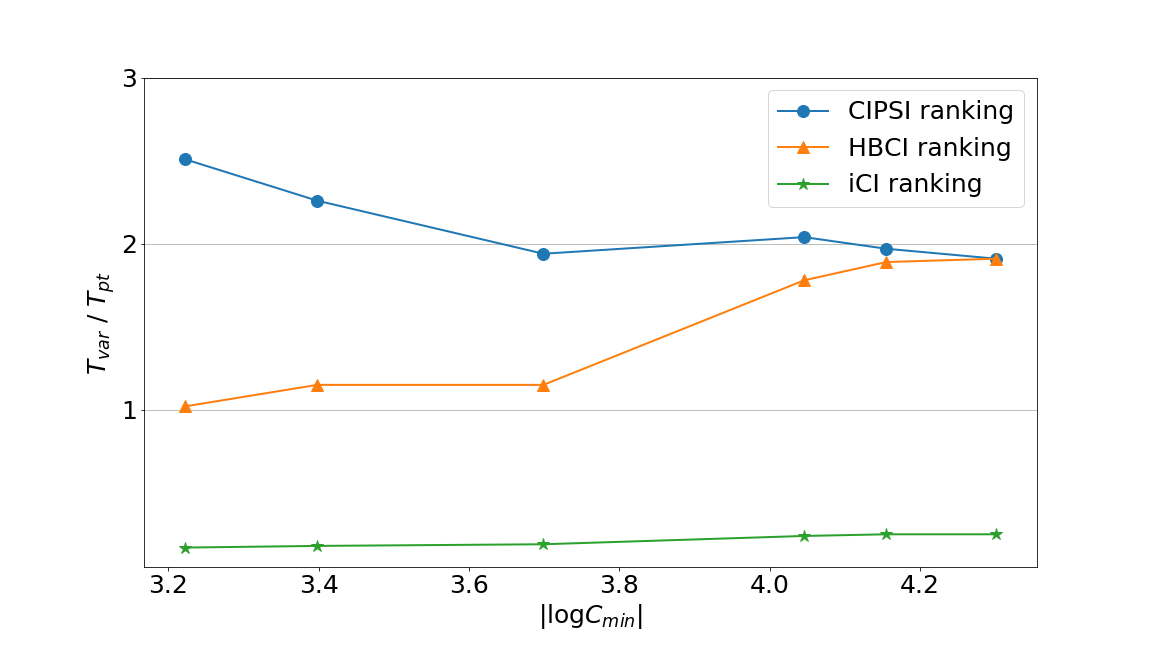}
	\caption{The time ratio $T_{var}/T_{pt}$ between selection and PT2 
by the CIPSI \eqref{CIPSIrank},
iCI \eqref{iCI_Rank_1}/\eqref{iCI_Rank_2} and HBCI \eqref{HBCI_Rank_1}/\eqref{HBCI_Rank_2} ranking criteria. }
	\label{TimeRatio}
\end{figure}

The very first point to be checked is the compactness of the final variational space $P$ determined by different ranking criteria discussed in Sec. \ref{Selection}.
To this end, all-electron nonrelativistic calculations were performed on the ground state of \ce{Cr2} with the Ahlrichs SV basis set\cite{AhlrichsSV} at two interatomic distances,
$R_{eq}$ (=1.68 ~\AA) and $2.0R_{eq}$. It can be seen from Fig. \ref{CR2} that, as the most rigorous ranking criterion,
CIPSI \eqref{CIPSIrank} does lead to the smallest number of CSFs and hence the most compact wave function for the same variational energy (NB: although not documented here,
the original iCI criterion \eqref{iCIrank}/\eqref{iCI-J} is virtually identical with CIPSI \eqref{CIPSIrank}). It appears that HBCI \eqref{HBCI_Rank_1}/\eqref{HBCI_Rank_2}
follows closely CIPSI. However, as can be seen from Fig. \ref{SizeRatio}, this arises with a high price: the variational space $P_1$ determined by HBCI \eqref{HBCI_Rank_1}/\eqref{HBCI_Rank_2}
is actually very large and is only reduced (by a factor of more than 10!) by the pruning step.
That is, the HBCI ranking criterion is too loose, thereby bringing in too many unimportant CSFs, %only less than 10\% of which survive the pruning.
more than 90\% of which are pruned away. This is clearly a waste of time for Hamiltonian construction and diagonalization.
Note in passing that the HBCI ranking used here is different from the original one\cite{HBCI2016}, which
does not invoke any pruning. It is also clear that the iCI criterion \eqref{iCI_Rank_1}/\eqref{iCI_Rank_2} leads to the least compact variational space $P$
among the three considered ranking criteria.
However, the situation gets improved steadily by reducing $C_{min}$. More importantly, the iCI selection is much more efficient than the other two (cf. Fig. \ref{TimeRatio}): the cost of
the iCI selection is only 20\% of that of PT2, whereas the other two are approximately two times more expensive than their PT2's. Therefore,
the slight loss of compactness is overcompensated by the gain in efficiency. An additional point deserves to be mentioned here:
while typically less than 1\% of the CSFs in the CIPSI and iCI variational spaces have coefficients somewhat smaller in absolute value than $C_{min}$,
such CSFs can be up to 10\% in the pruned HBCI variational space.

\subsection{\ce{Cr2}}
Given the improved efficiency of iCIPT2, we are now ready to perform more accurate calculations on the ground state of \ce{Cr2},
a classic strongly correlated system for testing various methods\cite{FCIQMC2014,HBCI2017a,HBCI-Cr2,SHBCI2018,ASCI2020,ASCI2018PT2,DMRG2015,DMRG2018}.
To this end, the spin-free exact two-component (sf-X2C) relativistic Hamiltonian\cite{X2CSOC1,X2CSOC2} and the cc-pVDZ-DK basis\cite{ccpv-dk}
were adopted, along with an active space of (28e, 76o) that is composed of $4.8\times10^{27}$ CSFs. While the Ne-core orbitals are just HF orbitals, the NOs
were used for correlation, which were generated in two ways,
(A) `partially dynamic' (see point (4) in the end of Sec. \ref{Selection}) and (B) `static'. In the latter, the NOs were generated with $C_{min}=1.0\times10^{-4}$ and then used
for all smaller $C_{min}$'s. In this case, the variational space determined at a $C_{min}$ can directly be used as initial guess for a smaller $C_{min}$.
The results with the two sets of NOs are documented in Tables \ref{Cr2_A} and \ref{Cr2_B}, respectively.
It is of interest to see that the extrapolated value (-2099.9223(4) $E_h$) by scheme B is very close to the HBCI value
(-2099.9224(6) $E_h$)\cite{SHBCI2018}.% but is significantly lower (by 2.8 $mE_h$) than the DMRG one (-2099.9195(27) $E_h$)\cite{DMRG2018}.
However, it is believed that the extrapolated value (-2099.9240(2) $E_h$) by scheme A is more accurate (by $-1.7$ $mE_h$) than that by scheme (B)
because the partially dynamically generated NOs in the former are of better quality than the fixed ones in the latter.
It is for sure that the two sets of calculations would agree with each other completely by further reducing $C_{min}$, so as to sample even larger portions of the full Hilbert space and hence
minimize orbital rotation effects. For instance, in the all-electron nonrelativistic calculations on \ce{Cr2} with the SV basis,
the difference between the energies
by scheme A (-2086.44468(2) $E_h$) and scheme B (-2086.44466(2) $E_h$)
is only $-0.02$ $mE_h$. Here, the largest FOIS ($1.3\times10^{11}$) sampled by iCIPT2/SV is ca. $10^{-8}$\% of the full space ($1.4\times10^{21}$).
In contrast, the largest FOIS ($3.2\times10^{12}$) sampled by sf-X2C-iCIPT2/cc-pVDZ-DK is only $10^{-14}\%$ of the
full space ($4.8\times10^{27}$).

\begin{table}[!htp]
	\small
	\centering
	\caption{Frozen-core sf-X2C-iCIPT2/cc-pVDZ-DK calculations of the ground state of \ce{Cr2} at $R_{eq}=1.68$~\AA~$^a$}
	\begin{threeparttable}
		\centering
\begin{tabular}{c|cccccccc}\toprule
	C$_{\text{min}}$\tnote{b}
	&\multicolumn{1}{c}{$N_{cfg}$\tnote{c}}
	&\multicolumn{1}{c}{$N_{csf}$\tnote{d}}
	&\multicolumn{1}{c}{$\tilde{N}_{csf}$\tnote{e}}
	&\multicolumn{1}{c}{$\tilde{N}_{det}$\tnote{f}}
	&\multicolumn{1}{c}{$E_{var}(+2099)/E_h$}
	&\multicolumn{1}{c}{$E_{tot}(+2099)/E_h$}
	&\multicolumn{1}{c}{$T_{wall}/s$}\\\toprule
	$5.0\times10^{-5}$&   455404 &  3136853 &  896527 &   3666946&  -0.81095 & -0.90466 &  464 \\
	$4.0\times10^{-5}$&   615595 &  4512684 & 1250882 &   5197093&  -0.81929 & -0.90612 &  641 \\
	$3.0\times10^{-5}$&   903347 &  7115256 & 1904606 &   8059852&  -0.82909 & -0.90785 &  973 \\
	$2.0\times10^{-5}$&  1560133 & 13369034 & 3470331 &  15033913&  -0.84161 & -0.90100 & 1775 \\
	$1.5\times10^{-5}$&  2281773 & 20737948 & 5268475 &  23176915&  -0.84926 & -0.91128 & 2765 \\
	$1.0\times10^{-5}$&  3959566 & 38959372 & 9616690 &  43154698&  -0.85945 & -0.91300 & 5194 \\
	$9.0\times10^{-6}$&  4565597 & 45851900 &11235054 &  50662572&  -0.86191 & -0.91341 & 6309 \\\midrule
	0.0\tnote{g}
	&&&&\multicolumn{3}{c}{-2099.9240$\pm$0.0002}&\\\bottomrule
\end{tabular}
\begin{tablenotes}
\item[a] D$_{2h}$ symmetry; HF energy: -2098.536329 $E_h$; active space: (28e, 76o); partially dynamically generated NOs (see point (4) in the end of Sec. \ref{Selection}).
\item[b] Threshold for pruning CSF in the variational (var) space.
\item[c] Number of orbital configurations in the variational space.
\item[d] Number of CSFs corresponding to $N_{cfg}$.
\item[e] Number of CSFs after selection, among which about 0.5\% have coefficients slightly smaller in absolute values than $C_{\mathrm{min}}$ (which is due to a final diagonalization).
\item[f]Estimated number of determinants according to the expression $\sum_I\frac{\tilde{N}_{\mathrm{csf}}^I}{N_{\mathrm{csf}}^I}N_{\mathrm{det}}^I$, with $N_{\mathrm{det}}^I$
being the numbers of determinants of CFG $|I\rangle$.
\item[g]Extrapolated value by linear fit of the $E_{\mathrm{total}}$ vs. $|E_c^{(2)}|$ plot, with uncertainty being half the length of 95\% confidence interval.
\end{tablenotes}
	\end{threeparttable}\label{Cr2_A}
\end{table}

\begin{table}[!htp]
	\small
	\centering
	\caption{Frozen-core sf-X2C-iCIPT2/cc-pVDZ-DK calculations of the ground state of
\ce{Cr2} at $R_{eq}=1.68$~\AA~$^a$}
	\begin{threeparttable}
		\centering
\begin{tabular}{c|ccccccc}\toprule
	C$_{\text{min}}$
	&\multicolumn{1}{c}{$N_{cfg}$}
	&\multicolumn{1}{c}{$N_{csf}$}
	&\multicolumn{1}{c}{$\tilde{N}_{csf}$}
	&\multicolumn{1}{c}{$\tilde{N}_{det}$}
	&\multicolumn{1}{c}{$E_{var}(+2099)/E_h$}
	&\multicolumn{1}{c}{$E_{tot}(+2099)/E_h$}
	&\multicolumn{1}{c}{$T_{wall}/s$}\\\toprule
	5.0$\times10^{-5}$ &  425431 &  2892927 &   840568 &  3438022&  -0.809883 & -0.903447 &434\\
	4.0$\times10^{-5}$ &  573307 &  4145556 &  1168810 &  4857157&  -0.818110 & -0.904792 &599\\
	3.0$\times10^{-5}$ &  840891 &  6513720 &  1779100 &  7530355&  -0.827922 & -0.906404 &916\\
	2.0$\times10^{-5}$ & 1449299 & 12182405 &  3233090 & 14006309&  -0.840470 & -0.908481 &1651\\
	1.5$\times10^{-5}$ & 2153481 & 19286806 &  4989382 & 21922383&  -0.847846 & -0.909859 &2645\\
	1.0$\times10^{-5}$ & 3716908 & 35893887 &  9034413 & 40531897&  -0.858790 & -0.911610 &4978\\
	9.0$\times10^{-6}$ & 4284579 & 42309173 & 10542471 & 47528961&  -0.861208 & -0.912028 &5608\\ \midrule
	0.0&&&&\multicolumn{3}{c}{-2099.9223$\pm$0.0004}&\\
	0.0\tnote{b}&&&&\multicolumn{3}{c}{-2099.9224$\pm$0.0006}&\\
	0.0\tnote{c}&&&&\multicolumn{3}{c}{-2099.9195$\pm$0.0027}&\\\bottomrule
\end{tabular}
\begin{tablenotes}
\item[a] Fixed NOs generated with $C_{min}=1.0\times 10^{-4}$. For additional explanations see Table \ref{Cr2_A}.
\item[b] HBCI\cite{SHBCI2018}.
\item[c] DMRG ($M=16000$)\cite{DMRG2018}.
\end{tablenotes}
	\end{threeparttable}\label{Cr2_B}
\end{table}

\subsection{[2Fe-2S]}
As a final example, we consider \ce{[Fe2S2(SCH3)]4}$^{2-}$ (abbreviated as [2Fe-2S]). Albeit a simplest system for
modeling iron-containing enzymes\cite{spDMRG2017}, even the minimal chemically meaningful
active space (30e, 20o) already contains  $240374016$ CSFs ($52581816$, $99419400$, $64438500$, $20575100$, $3174444$, and $184756$ CSFs for spins from 0 to 5, respectively).
The 20 active orbitals here include \ce{Fe} $3d$ and \ce{S} $3p$ of the core [2Fe-2S]
as well as their four $\sigma$-bonds with ligands.
To make a direct comparison with the previous work, the (nonrelativistic) molecular orbital integrals are taken simply from Ref. \citenum{spDMRG2017}.
The extrapolated (valence) energy of the singlet ground state is -116.60574(6) $E_h$ (cf. Table \ref{FeS_Singlet}),
very close to the DMRG value of -116.60561 $E_h$ with $M=8000$\cite{LiFe2S2}.
It is more appealing to take a closer look at the wave function (selected with
$C_{min}=5.0\times10^{-6}$): it is composed of 41.13\%  \ce{Fe}(II) -- \ce{Fe}(II), 22.69\% \ce{Fe}(III) -- \ce{Fe}(II),
22.01\% \ce{Fe}(II) -- \ce{Fe}(I), 6.05\% \ce{Fe}(III) -- \ce{Fe}(I), 3.40\% \ce{Fe}(III) -- \ce{Fe}(III), and 2.86\% \ce{Fe}(I) -- \ce{Fe}(I),
but the leading CSF has a weight only of 3.36\% and stems from  \ce{Fe}(III) -- \ce{Fe}(III) instead of the leading structure \ce{Fe}(II) -- \ce{Fe}(II).
It is not clear whether this peculiar picture will hold for a larger active space, especially in conjunction with localized orbitals.
Nevertheless, it is perfectly legitimate to further calculate more spin states, just to reveal the efficacy of iCIPT2.

\begin{figure}[!htp]
	\centering
	\includegraphics[width=0.7\textwidth]{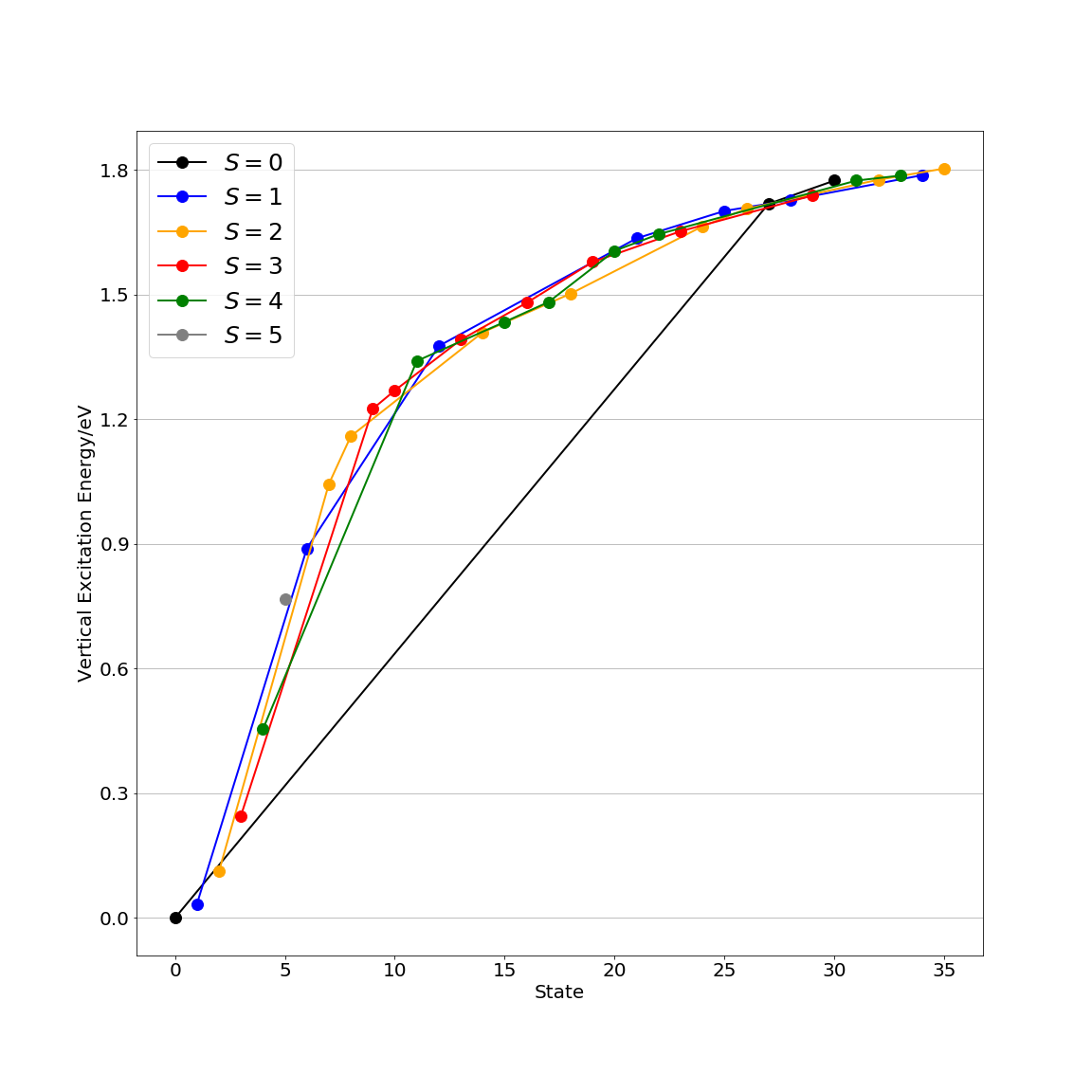}
	\caption{Vertical excitation energies of low-lying excited states of [2Fe-2S]}
	\label{FeSspectrum}
\end{figure}

Although the particle-hole algorithm presented in Sec. \ref{FastHmat} allows for simultaneous calculations of several
states of different spins with a common set of orbitals, the function is not yet at our disposal. Instead, states of different spins are
calculated here separately. One then has to face the issue that
the quality of states of different spins may be different for the same $C_{min}$, so as to affect the relative energies.
To circumvent this problem, we decompose the vertical excitation energy $\Delta E_{S,i}$ ($=E_{S,i}-E_{0,0}$) of state $i$ with spin $S$ into two terms,
$\Delta E_{S,i}-\Delta E_{S,0}$ and $\Delta E_{S,0}-\Delta E_{0,0}$. The former is calculated with the same $C_{min}$, whereas the latter is set to the extrapolated value, i.e.,
\begin{equation}
\Delta E_{S,i}(0.0) = \left[E_{S,i}(C_{min})-E_{S,0}(C_{min})\right]+\left[E_{S,0}(0.0)-E_{0,0}(0.0)\right].
\end{equation}
In essence, what is assumed here is that different states of the same spin have the same extrapolation distance.
The so-calculated excitation energies for all states below 1.8 eV are documented in Table \ref{FeSall} and further plotted in Fig. \ref{FeSspectrum}.
Interestingly, the lowest 6 states originate just from the lowest state of each of the 6 possible spins $S\in[0,5]$ that result from
the coupling of the two high-spin-$\frac{5}{2}$ ionic configurations Fe(III) $3d^5$ and are energetically ordered in increasing spin.
However, the remaining 30 states of different spins are intersected within an interval of 0.91 eV. That is, they are
separated in average by only 0.03 eV, just like the gap between the lowest singlet and triplet states.
It is clear that such complicated spin structure cannot be handled by determinant-based methods due to severe spin contaminations.

\begin{table}[!htp]
	\small
	\caption{The singlet ground state of \ce{[Fe2S2(SCH3)]4}$^{2-}$ (for additional explanations see Table \ref{Cr2_A})}
	\begin{threeparttable}
		\centering
		\begin{tabular}{c|cccccc}\toprule
			C$_{\text{min}}$
			&\multicolumn{1}{c}{$N_{cfg}$}
			&\multicolumn{1}{c}{$N_{csf}$}
			&\multicolumn{1}{c}{$\tilde{N}_{csf}$}
			&\multicolumn{1}{c}{$\tilde{N}_{det}$}
			&\multicolumn{1}{c}{$E_{var}/E_h$}
			&\multicolumn{1}{c}{$E_{tot}/mE_h$}
			\\\toprule
			$2.0\times10^{-5}$ &   313428 &  6242941 &  568382 &  3056319& -116.602301 &  -116.604325 \\
			$1.5\times10^{-5}$ &   396163 &  7436740 &  754898 &  4031212& -116.603153 &  -116.604691 \\
			$1.0\times10^{-5}$ &   540358 &  9332213 & 1108867 &  5866360& -116.604010 &  -116.605046 \\
			$9.0\times10^{-6}$ &   586437 &  9908183 & 1228988 &  6485715& -116.604196 &  -116.605120 \\
			$7.0\times10^{-6}$ &   744932 & 11779934 & 1673904 &  8765188& -116.604554 &  -116.605256 \\
			$5.0\times10^{-6}$ &   938275 & 13485301 & 2236247 & 11619598& -116.604908 &  -116.605385 \\\bottomrule
			0.0
			&&&&&\multicolumn{2}{c}{-116.60574$\pm$0.00006}\\\bottomrule
		\end{tabular}
	\end{threeparttable}\label{FeS_Singlet}
\end{table}

\begin{table}
	\caption{Vertical excitation energies ($\Delta E$ in eV) of low-lying excited states of [2Fe-2S]}
	\begin{tabular}{ccc|ccc}\toprule
		State & $2S+1$ & $\Delta E$ &
		State & $2S+1$ & $\Delta E$ \\\toprule
		0   & 1 & 0.000 & 18  & 5 & 1.502 \\
		1   & 3 & 0.034 & 19  & 7 & 1.578 \\
		2   & 5 & 0.112 & 20  & 9 & 1.605 \\
		3   & 7 & 0.246 & 21  & 3 & 1.635 \\
		4   & 9 & 0.454 & 22  & 9 & 1.645 \\
		5   &11 & 0.767 & 23  & 7 & 1.652 \\
		6   & 3 & 0.889 & 24  & 5 & 1.664 \\
		7   & 5 & 1.044 & 25  & 3 & 1.701 \\
		8   & 5 & 1.159 & 26  & 5 & 1.706 \\
		9   & 7 & 1.225 & 27  & 1 & 1.718 \\
		10  & 7 & 1.269 & 28  & 3 & 1.727 \\
		11  & 9 & 1.340 & 29  & 7 & 1.738 \\
		12  & 3 & 1.376 & 30  & 1 & 1.774 \\
		13  & 7 & 1.391 & 31  & 9 & 1.774 \\
		14  & 5 & 1.409 & 32  & 5 & 1.776 \\
		15  & 9 & 1.434 & 33  & 9 & 1.786 \\
		16  & 7 & 1.480 & 34  & 3 & 1.787 \\
		17  & 9 & 1.482 & 35  & 5 & 1.803 \\ \bottomrule
	\end{tabular}\label{FeSall}
\end{table}

%\begin{table}[!htp]
%	\small
%	\caption{The ground state of states with different spin multiplicity $2S+1$}
%	\begin{threeparttable}
%		\centering
%		\begin{tabular}{cccc}\toprule
%		Spin Multiplicity & FCI dimension (CSF) & Energy\tnote{a} /$E_h$ & Gap\tnote{b} /eV \\\toprule
%		1                 &52581816& -116.60574 & -\\
%		3                 &99419400& -116.60448 & 0.034\\
%		5                 &64438500& -116.60161 & 0.112\\
%		7                 &20575100& -116.59669 & 0.246\\
%		9                 &3174444 & -116.58904 & 0.454\\
%		11                &184756  & -116.57754 & 0.767\\\bottomrule
%		\end{tabular}
%	\begin{tablenotes}
%		\item[a] The extrapolated energy is obtained by linear extrapolating $E_{tot}$ versus $E_{pt}$ with $C_{min}\in\{2.0,1.5,1.0,0.9,0.7,0.5\}\times10^{-5}$. The extrapolationg error of all the six states above is within 0.1 m$E_h$.
%		\item[b] Energy deviation from the ground state of singlet states.
%	\end{tablenotes}
%	\end{threeparttable}\label{FeS_Spin}
%\end{table}

\section{Conclusions and Outlook}\label{Conclusion}
Sticking to the parlance of static and dynamic correlations, there could be two paradigms for handling strongly
correlated systems of electrons, viz. ``more static, less dynamic'' and ``less static, more dynamic''.
It is the former that is followed by the family of sCI+PT2 methods, among which iCIPT2 stands out in several aspects as already highlighted in the Introduction.
Its efficiency has been improved herein by up to $20\times$ via three major techniques: a new criterion for configuration selection, a new particle-hole
algorithm for Hamiltonian construction, and a new data structure for the quick sorting of the FOIS, 
in addition to the workhorse TUGA for computing and reutilizing the basic coupling coefficients between randomly selected CSFs. 
All these can be applied to other types of sCI+PT2 as well.
As revealed by several examples, iCIPT2 can indeed be characterized as an near-exact approach. However, it is still memory intensive, even though
the FOIS has been decomposed into disjoint subspaces. A possible way to resolve this issue is to evaluate the PT2 correction in a stochastic manner.
Other immediate extensions of iCIPT2 include simultaneous treatment of states of different spatial and/or spin symmetries (which is furnished by
the particle-hole algorithm), perturbative treatment of spin-orbit couplings (via the sf-X2C+soc-DKH1 Hamiltonian\cite{X2CSOC1,X2CSOC2}), and direct access
of high-lying states of a given energy window (which is furnished by the iVI eigensolver\cite{iVI,iVI-TDDFT}), etc.
Work along these directions are being undertaken at our laboratory.

\begin{acknowledgement}
The research of this work was supported by National Natural Science Foundation of China (Grant Nos. 21833001 and 21973054)
and the North Dakota University System.
\end{acknowledgement}

\bibliography{iCI}

\iffalse

\newpage
For TOC only

\includegraphics[width=\textwidth]{TOC}

\fi

\end{document}